\documentclass[doublecolumn]{mn2e}
\usepackage{epsfig}
\newcommand{\mbh}{M_{\rm{BH}}}
\newcommand{\msat}{M_{\rm{sat}}}
\newcommand{\mhost}{M_{\rm{host}}}

\newcommand{\rvir}{r_{\rm{vir}}}

\newcommand{\tdf}{\tau_{\rm merge}}

\def\kms{\>{\rm km}\,{\rm s}^{-1}}

\def\Msun{\>{\rm M_{\odot}}}

\def\Lsun{\>{\rm L_{\odot}}}

\newcommand{\gtsim}{\mathrel{\hbox{\rlap{\lower.55ex \hbox {$\sim$}}
                   \kern-.3em \raise.4ex \hbox{$>$}}}}
\newcommand{\ltsim}{\mathrel{\hbox{\rlap{\lower.55ex \hbox {$\sim$}}
                   \kern-.3em \raise.4ex \hbox{$<$}}}}

\title[Local Massive Black Holes]{Growing Massive Black Holes in a Local Group Environment: the Central Supermassive, Slowly Sinking, and Ejected Populations }

\author[Miroslav Micic, Kelly Holley-Bockelmann, Steinn Sigurdsson]
{Miroslav Micic$^1$\thanks{E-mail: m.micic@physics.usyd.edu.au}, Kelly Holley-Bockelmann$^{2,3}$, Steinn Sigurdsson$^4$ \\
$^1$ Sydney Institute for Astronomy, School of Physics, The University of Sydney \\
$^2$ Department of Physics \& Astronomy, Vanderbilt University  \\
$^3$ Department of Physics, Fisk University \\
$^4$ Department of Astronomy \& Astrophysics, Pennsylvania State University  \\
}

\begin{document}
\maketitle

\begin{abstract}

We explore the growth of  $\leq$10$^7\Msun$ black holes that reside at the centers 
of spiral and field dwarf galaxies in a Local Group type of environment. We use merger 
trees from a cosmological N-body simulation known as Via Lactea II (VL-2) as a 
framework to test two merger-driven semi-analytic recipes for black hole growth that 
include dynamical friction, tidal stripping, and gravitational wave recoil in over 
20,000 merger tree realizations. First, we apply a Fundamental Plane limited (FPL) 
model to the growth of Sgr A*, which drives the central black hole to a maximum mass 
limited by the Black Hole Fundamental Plane after every merger. Next, we present a 
new model that allows for low-level Prolonged Gas Accretion (PGA) during the merger. 
We find that both models can generate a Sgr A* mass black hole.  We predict a population 
of massive black holes in local field dwarf galaxies -- if the VL-2 simulation is representative 
of the growth of the Local Group, we predict up to 35 massive black holes ($\leq$ 10$^6\Msun$) 
in Local Group field dwarfs. We also predict that hundreds of $\leq$ 10$^5\Msun$ black holes fail to 
merge, and instead populate the Milky Way halo, with the most massive of them at roughly the 
virial radius. In addition, we find that there may be hundreds of massive black holes 
ejected from their hosts into the nearby intergalactic medium due to gravitational 
wave recoil. We discuss how the black hole population in the Local Group field dwarfs may 
help to constrain the growth mechanism for Sgr A*.
\end{abstract}


\section{INTRODUCTION}

Supermassive black holes, with masses of $10^6 {\rm M}_\odot \leq {\rm M} \leq 10^{10} {\rm M}_\odot$, 
are widely believed to dwell at the centers of elliptical galaxies and spiral bulges (e.g. Kormendy 
$\&$ Richstone 1995); the best known example is observed at the center of the Milky Way, with a mass 
M$_{\rm SMBH}$ = 4.2$\times10^6\Msun$ (Ghez et al. 2008). There is abundant evidence that when a SMBH 
is in place, it transforms the structure and evolution of the galaxy, from powering active galactic 
nuclei at high redshifts (Greenstein $\&$ Matthews 1963, Rees 1984, Alexander et al. 2005, Fan 2005), 
to regulating star formation throughout the galaxy (Di Matteo et al. 2005, Croton et al. 2006, 
Cox et al. 2008), to scouring the galactic nucleus of stars during SMBH mergers (Ebisuzaki et al. 1991, 
Quinlan 1996, Makino 1997, Milosavljevic $\&$ Merritt 2001, Volonteri et al. 2003).

This deep connection between the evolution of SMBHs and galaxies is perhaps best encapsulated in a 
remarkable correlation between the SMBH mass and the velocity dispersion of the host spheroid (Gebhardt 
et al. 2000, Ferrarese $\&$ Merritt 2000, Tremaine et al. 2002, Marconi $\&$ Hunt 2003, Hopkins et al. 
2007). The dispersion in the  {\em black hole fundamental plane} (Hopkins et al. 2007) points to an 
intrinsically tight correlation, at least for a sample of nearby bright spiral and elliptical galaxies 
with clear dynamical SMBH signatures. However, on a smaller mass scale, in systems that have a mass 
comparable to the Milky Way mass or smaller, central SMBHs may become less common as bulges become less 
common (Ferrarese et al. 2006), and in some instances they disappear entirely, such as in the case of 
M33 (Merritt et al. 2001, Gebhardt et al. 2001), and NGC 205 (Valluri et al. 2005). Many of these 
bulgeless stellar systems host a nuclear star cluster instead. Nuclear star clusters are found in 
late-type spirals (Boker et al. 2002), and dwarf elliptical galaxies (Cote et al. 2006). This changeover 
may be the result of a competition between the SMBH and nuclear star cluster for the same gas reservoir. 
Nayakshin et al.(2009) shows that in massive galaxies with a spheroid velocity dispersion of 
$\sigma$ $\geq$ 150 $\kms$, gas accretion onto the black hole dominates and a SMBH forms, while in 
galaxies with lower velocity dispersions, star formation generates a nuclear cluster. Note that such a 
star cluster does not exclude presence of an underweight SMBH (Nayakshin et al. 2009).

Several observational and theoretical studies have linked the SMBH mass to the mass of the 
host dark matter halo (Ferrarese 2002, Baes et al. 2003, Shankar et al. 2006). This 
relation is a boon to theorists because many of the leading explanations of SMBH 
birth and growth are driven by hierarchical structure formation (Hopkins et al. 2005, 
Wyithe $\&$ Loeb 2005, Granato et al. 2001, Menou et al. 2001, Adams et al. 2001, 
Monaco et al. 2000, Silk $\&$ Rees 1998, Haehnelt et al. 1998, Haehnelt $\&$ Kauffmann 
2000, Cattaneo et al. 1999, Loeb $\&$ Rasio 1994), and are therefore tied to the mass 
of the dark matter halo.

In the current picture of SMBH assembly, the black hole begins life as a low mass 
``seed'' black hole at high redshift. It's not clear, though, when exactly these 
BH seeds emerge or what mass they have at birth. SMBH seeds may have been spawned 
from the accretion of low angular momentum gas in a dark matter halo (Koushiappas 
et al. 2004, Bromm $\&$ Loeb 2003, 2004), the coalescence of many seed black holes within 
a halo (Begelman $\&$ Rees 1978, Islam et al. 2004), or from an IMBH formed, perhaps, 
by runaway stellar collisions (Portegies Zwart et al. 2004, Miller $\&$ Colbert 2004, 
van der Marel 2004) or they could even be primordial (Mack et al. 2007). However, 
the most likely candidates for SMBH seeds are the remnants that form from the first 
generation of stars sitting deep within dark matter halos (Madau $\&$ Rees 2001, 
Heger et al. 2003, Volonteri et al. 2003, Islam et al. 2003, Wise $\&$ Abel 2005) 
-- so called Population III stars. With masses less than roughly $ 10^3 \Msun$, these relic seeds are 
predicted to lie near the centers of dark matter halos at high redshifts (Bromm 
et al. 1999, Abel et al. 2000, 2002). Structure formation dictates that dark matter 
halos form in the early universe and hierarchically merge into larger bound objects, 
so naturally as dark matter halos merge, seed black holes sink to the center through 
dynamical friction and eventually coalesce.

Gas accretion is thought to play a critical role in fueling the early stages of black 
hole growth (David et al. 1987, Kauffmann $\&$ Haehnelt 2000, Merloni 2004), and this 
may explain the tightness of the M$_{\rm BH}-\sigma$ relation (Burkert $\&$ Silk 2001,
Haehnelt $\&$ Kauffmann 2000, Di Matteo et al. 2005, Kazantzidis et al. 2005, Robertson 
et al. 2006). Since high redshift galaxies are thought to be especially gas-rich, each 
merger brings a fresh supply of gas to the center of the galaxy, and new fuel to the 
growing supermassive black hole (Mihos $\&$ Hernquist 1994, Di Matteo et al. 2003). From 
a combination of gas accretion and binary black hole coalescence, it is thought that these 
Pop III-generated seeds may form the SMBHs we observe today (Soltan 1982, Schneider et al. 2002).

During a galaxy merger, each black hole sinks to the center of the new galaxy potential 
due to dynamical friction and eventually becomes bound as a binary (Kazantzidis et al. 
2005; Escala et al. 2005). Dynamical friction then continues to shrink the orbit until 
the binary is hard (i.e, the separation between each black hole, a$_{\rm BBH}$, is such 
that the system tends to lose energy during stellar encounters) (Heggie et al. 2007).
Thereafter, further decay is mediated by 3-body scattering with the ambient stellar 
background until the binary becomes so close that the orbit can lose energy via gravitational 
radiation. In studies of static, spherical potentials, it may be difficult for stellar 
encounters alone to cause the binary to transition between the 3-body scattering phase
and the gravitational radiation regime (Milosavljevic $\&$ Merritt 2003). However, in 
gas-rich or non-spherical systems, the binary rapidly hardens and coalesces into one
black hole, emitting copious gravitational radiation in the process (Mayer et al. 2007, 
Kazantzidis et al. 2005, Berczik et al. 2006, Sigurdsson 2003, Holley-Bockelmann $\&$ Sigurdsson 2006).

Unless the black hole binaries stall at the final parsec, the longest timescale governing 
the coalescence of two black holes occurs when the host galaxies themselves are still merging. 
Here, the dynamical evolution of two merging galaxies is driven by the combined effect of 
{\em dynamical friction}, which brings the less massive galaxy, or satellite, to the center of 
the larger halo, or primary, and {\em tidal interaction} which strips mass from the satellite 
and further delays the merger (e.g. Richstone 1976, Aguilar $\&$ White 1986, Holley-Bockelmann 
$\&$ Richstone 1999, Taffoni et al. 2003). If the dynamical friction timescale is longer than a 
Hubble time, the black holes carried by their host galaxies will not sink close enough to form a binary.

Sijacki et al. 2007 and Di Matteo et al. 2008 have performed state-of-the-art high resolution 
hydrodynamic simulations of cosmological structure formation, following the growth of high mass 
SMBHs at the centers of massive elliptical galaxies and clusters of galaxies. Their research was 
followed by similar semi-analytic work that incorporates a full treatment of dark matter dynamics, 
radiative gas cooling, star formation and energy feedback processes (Somerville et al. 2008). 
In this very elegant approach, the SMBH accretes gas through a {\em quasar mode} -- nearly Eddington 
rate accretion following a galaxy merger (Croton et al. 2006) -- and a {\em radio mode } -- Bondi-Hoyle 
accretion associated with relativistic jets (Somerville et al. 2008).  Both modes produce feedback 
that heats the surrounding gas. In this model, the feedback stops the accretion and locks the growth 
of the SMBHs to the fundamental plane. At the same time, the feedback also quenches the star formation, 
which explains the observed shallow metallicity, stellar density and entropy profiles. However, it has 
recently been suggested that the implementation and importance of AGN feedback may need to be reexamined 
(Ostriker et al. 2010). Also, radiation fields and winds produced by massive stars may provide the 
dominant feedback (Hopkins et al. 2010). In fact, it is worth mentioning that AGN feedback is just one 
of many feedback processes occurring in galaxy centers; supernovae, star formation, and galaxy mergers 
produce feedback as well (Sinha $\&$ Holley-Bockelmann 2010). It is not clear which feedback process 
contributes the most to the evolution of a given galaxy.

As well-developed as the current effort to understand SMBH growth is, most work has focused on growing 
the most massive SMBHs, found predominantly in dense environments. In a Local Group environment though, 
the galaxy morphology, dynamics, and star formation history are all dramatically different. Smaller 
stellar systems such as disk and dwarf galaxies, have SMBHs with lower masses ($\leq$ 10$^6\Msun$) 
or no SMBHs at all for bulgeless galaxies like M33. In cosmological simulations, AGN feedback models 
create significant uncertainties for black hole growth in these lower mass halos (Booth $\&$ Schaye 2009). 
In addition, SMBHs in the Local Group are currently subluminous, implying a different relationship 
with gas consumption, at least in the present epoch, than the more massive end of the SMBH mass spectrum.
Nonetheless, mergers are still expected to provide a key mechanism to channel gas to the galactic center.  
While ``cold mode''   gas accretion (Keres et al. 2005, Dekel et al. 2009) can channel large amounts 
of cold gas ($\leq 10^4$ K ) to the galactic disk, this gas may not fuel the SMBH.  Indeed, high resolution 
N-body-SPH simulations of Milky Way formation show that most of the gas that is accreted originates from 
major mergers; even during the ``radio mode'', where the gas is accreted by a Bondi - Hoyle mechanism,  
the gas reservoir is stocked not from the cold mode, but from the major mergers the Milky Way finished 
billions of years ago (J. Bellovary, private communication). Hence, galaxy mergers appear to drive SMBH 
growth in a wide range of masses. 

One problem that can preferentially plague low mass galaxies is gravitational wave recoil. Recent 
calculations of binary black hole mergers indicate that gravitational wave recoil can kick a newly 
merged black hole with a speed as large as $\sim$4000$\kms$ (Herrmann et al. 2007, Gonzalez et al. 2007a, 
2007b, Koppitz et al. 2007, Campanelli et al. 2007, Schnittman $\&$ Buonanno 2007). The magnitude of 
the gravitational wave recoil imparted during any given merger depends on the mass ratio of each black 
hole, the spin magnitude and orientation with respect to the binary orbital plane, as well as the 
eccentricity of the orbit (Campanelli et al. 2007, Schnittman $\&$ Buonanno 2007, Baker et al. 2007). 
While certain spin alignment mechanisms (Bogdanovic et al. 2007, Dotti et al. 2010) could allow massive 
galaxies to retain their SMBHs, even moderate kicks can eject a growing SMBH from dwarf or high redshift 
galaxies (Merritt et al. 2004, Micic et al. 2006, Volonteri 2007, Schnittman 2007, Sesana 2007, Volonteri 
et al. 2010). There is tentative observations that suggest that gravitational wave recoil can indeed 
eject a SMBH from its host (Komossa et al. 2008). Given the greater vulnerability to recoil, the 
questionable importance of AGN feedback, and lack of SMBH in some galaxies, one can argue that  SMBH 
growth may be substantially different for these lightest SMBHs in Local Group type of environments.

In this paper we use Via Lactea 2 (VL-2) evolutionary tracks to begin with an N-body merger tree that 
mimics the assembly of the Milky Way halo (Diemand et al. 2007,2008). Upon this numerical merger tree we 
paint semi-analytic recipes for the galaxy and black hole growth. We follow two different merger driven
recipes. The first approach is inspired by recent semi-analytic and numerical work on black hole growth 
(Somerville et al. 2008, Croton et al. 2006, Hopkins et al. 2005, 2007). Here, SMBHs grow slowly 
through low-level Bondi-Hoyle gas accretion tempered by AGN feedback -- the radio mode; this growth is 
punctuated by rapid Eddington gas accretion triggered by a major merger -- the quasar mode. This approach 
has been used to explain SMBH growth in massive stellar systems but not the growth of massive black 
holes in the Local Group. Here, SMBH growth is limited by the black hole fundamental plane. Hence, 
hereafter we call this approach Fundamental Plane limited (FPL). In the second recipe, we introduce a 
model for prolonged black hole gas accretion physically motivated by galaxy dynamics. In this model, 
the subgrid physics (stellar and AGN feedback, as well as accretion disk microphysics) is bundled into 
one parameter for gas accretion efficiency that can be constrained by future small-scale, fully general 
relativistic magneto-hydrodynamic AGN simulations. With this approach, gas accretion onto the black hole 
is based on the dynamics of the galaxy merger -- fast at high rate for major mergers and prolonged 
at low rate for more minor mergers. This model does not use the black hole fundamental plane or the 
M - $\sigma$ relation to limit black hole growth. During a minor merger, the accretion time scale is 
longer, but the accretion rate is lower, since less gas is driven toward the BH by a smaller global 
perturbation. For major mergers, the accretion rate increases, though in practice it never reaches the 
Eddington rate. We call this approach Prolonged Gas Accretion (PGA). In addition to gas accretion, black 
holes grow through mergers with other black holes too. In both recipes, we model SMBH growth from direct 
mergers including gravitational recoil. Since the recoil velocity is highly dependent on BH spin and 
orientation, we create $\sim$ 20,000 merger trees to examine the combination of BH spin parameters that 
favors the $M$ -- $\sigma$ relation at redshift zero.  

The key difference between FPL and PGA models is in the implementation of AGN feedback. 
In FPL, the black hole fundamental plane is used to calibrate AGN feedback. In PGA, we 
use AGN feedback to suppresses the efficiency of gas accretion, while calibration 
parameters come from detailed small scale simulations of galaxy merger remnants.

We describe our method in section 2 and introduce our black hole growth prescriptions. 
In section 3, we present results, focusing on the evolution of our Sgr A* analogue in the
FPL and PGA models. In Section 4, we present predictions for the local population of 
massive black holes, rogue black holes in the Milky Way, and ejected black holes into 
the nearby intergalactic medium. We discuss the implications of our results and future 
work in section 5.

\section{METHOD}

\subsection{VL-2 Dark Matter Halo Merger Tree}

In this paper, we use publicly-available VL-2 evolutionary tracks (Diemand et al. 
2007,2008). VL-2 is the highest resolution cosmological N-body simulation of Milky 
Way formation and evolution, and evolves a $\Lambda$CDM universe with WMAP3 parameters 
($\Omega_{\rm M}$=0.238, $\Omega_{\Lambda}$=0.762, $\sigma_8$=0.74 and h=0.7). 
The high resolution region of VL-2 is embedded within a periodic box with a comoving 
length of 40 Mpc. The evolutionary tracks consist of the time evolution from z=27.54 to 
z=0 of $\sim$ 20,000 dark matter halos identified at redshift z=4.56. Halos have masses 
as small as $\sim$ 10$^5\Msun$.

In our hybrid method, we combine dark matter halo merger trees obtained from VL-2 
evolutionary tracks with an analytical treatment of the physical processes that arise in the 
dynamics of galaxy and black hole mergers. Our N-body approach stops with the creation 
of the halo merger tree. We seed dark matter halos with Population III black holes until z=5, 
following the work of Trenti $\&$ Stiavelli 2009 (Fig 1 in Trenti et al. 2009), and follow their 
merger history from redshift 27.54 to 0 by constructing numerical merger trees interpolating 
between the snapshots at z$\leq$ 15. To define the structure of each dark matter halo within the 
N-body generated merger tree, we assume a Navarro, Frenk, $\&$ White (hereafter, NFW) 
density profile (Navarro, Frenk $\&$ White 1996). We set the parameters of a given NFW 
halo using the approach presented in Bullock et al. 2001, assuming the typical virial 
mass of a dark matter halo to be M$_{\rm typ}$ = 1.5 $\times$ 10$^{13}$$\Msun$ at 
redshift zero. Note that only halos that are seeded with black holes or merging with other halos 
are retained in our merger tree. For each merger, we tag the more massive halo as the primary, 
with mass M$_{\rm p}$, and the less massive halo as the secondary or satellite halo with mass 
M$_{\rm s}$. Note that the properties of low mass dark matter halos in the mass range of VL-2 
(10$^{5}\Msun$ $\leq$ M $\leq$ 2$\times$10$^{12}$ $\Msun$) have not been studied in detail at z $\geq$ 3. 

Note that VL-2 evolutionary tracks do not include any halos or subhalos that were not present 
at z=4.56. Some halos are completely disrupted in mergers prior to that time (no subhalo counterpart 
at z=4.56), and some halos form after z=4.56. These halos are small in mass and would have been cut 
out of the merger tree anyway since they are either too small to host the black hole seed or too 
small to have relevant amounts of cold gas for either black hole accretion or star formation.
Note that VL-2 does not necessarily represent the actual merger histories of Milky Way or Local Group.

\subsection{SMBH Growth Prescriptions}

We test two fundamentally different models for SMBH growth: Fundamental Plane Limited (FPL) 
and Prolonged Gas Accretion (PGA). In FPL, gas accretion onto the black hole is controlled by 
AGN feedback in such way that the SMBH mass does not grow above the black hole fundamental 
plane at any time. In PGA, we bundle star formation and AGN feedback, as well as accretion 
disk microphysics together into gas accretion efficiency parameter; the mass is not restricted 
to lie on the black hole fundamental plane.

Our only guideline is the mass of Sgr A* observed today. Since the mass of the SMBH at the center 
of the Milky Way is well-constrained, we can identify which physical processes during the growth 
are most important for the final black hole mass and test their parameter space. The first of these 
parameters applies to both models: the black hole seed's initial mass function (BHMF).
We test various BHMFs and various values for the minimum and maximum black hole seed mass. 
A second effect that is present in both models is gravitational wave recoil (see Section 2.4).
The third parameter is unique in each model. In the FPL model, the third parameter is the mass 
accreted during the ``radio'' or quiescent mode of the AGN duty cycle. The SMBH growth during 
this phase is usually approximated by Bondi-Hoyle gas accretion, though this is not strictly necessary. 
We test various values for average mass accretion rate during this radio mode. Since the Bondi-Hoyle 
mechanism is commonly used, we may occasionally address this quiescent part of SMBH growth as the 
Bondi-Hoyle phase, but our approach does not depend on the actual growth mechanism here. In the PGA model, 
the third parameter is the gas accretion efficiency. In our implementation this parameter contains 
information on how the microphysics acts to suppress accretion onto the black hole, such as SNe feedback, 
accretion disk physics, etc. We treat all of these processes as free parameters to study the observable 
consequences at z=0.

In all models, we adopt Gnedin (2000) and Kravtsov et al. (2004) initial cold gas fractions 
(f$_{\rm cg}$ $\sim$ 10$^{-5}$ for M$_{\rm halo}$ $\sim$ 3$\times$10$^8\Msun$;          
f$_{\rm cg}$ $\sim$ 10$^{-2}$ for M$_{\rm halo}$ $\sim$ 4$\times$10$^9\Msun$; 
f$_{\rm cg}$ $\sim$ 0.13 for M$_{\rm halo}$ $\geq$ 1$\times$10$^{11}\Msun$)
in high redshift galaxies for a reionization range of 
10 $\leq$ z $\leq$ 11. In this manner we follow the approach of Somerville et al.(2008) for 
creating the gas reservoirs used for star formation and gas accretion onto the black holes. 
We do not include further multi mode accretion in the merger tree (i.e. Keres et al. 2005).

\subsubsection{Fundamental Plane Limited Model}

In the first model, FPL, we use the approach adopted by Somerville et al. (2008), but focus only 
on those aspects that are important for SMBH growth. In FPL, the SMBH grows through mergers and 
gas accretion. Gas accretion is controlled by two modes of AGN feedback. The quasar or ``bright'' 
mode occurs after the host galaxy goes through a major merger; this causes the central black hole 
to accrete gas exponentially until it reaches the Eddington limit, and it continues accreting at 
Eddington limit until it reaches a mass specified by the black hole fundamental plane. By design, 
the black hole mass at the end of each merger is proportional to the stellar mass of the spheroid 
M$_{\rm sph}$ in the model:

\begin{equation}
\log (\mbh/M_{\rm sph}) = -3.27 + 0.36 \, {\rm erf}[(f_{\rm gas}-0.4)/0.28],
\end{equation}
where $f_{\rm gas}$ is the amount of cold gas available for accretion (Somerville et al. 2008).
The quasar mode is replaced by a radio mode until the next major merger. During the radio mode, 
the black hole accretes gas at lower rates associated with radiatively inefficient Bondi-Hoyle 
accretion, such as what occurs in an advection dominated accretion flow model (Fender et al. 2004).
The Bondi-Hoyle accretion rate controls the black hole mass during the longer quiescent phase between 
galaxy mergers. In the Milky Way today, the SMBH mass is $4.2\times 10^6 \Msun$ (Ghez et al. 2008), and 
the Bondi-Hoyle rate is $\sim 5 \times 10^{-5}\Msun {\rm yr}^{-1}$ (Quataert $\&$ Gruzinov 2000, 
Melia $\&$ Falcke 2001). Since we know the current Sgr A* mass and the accretion rate observed today, 
we can determine how well the FPL model can match observations. To simplify our investigation we adopt 
an average Bondi-Hoyle accretion rate during the radio mode over a Hubble time. The Bondi-Hoyle accretion 
rate in our model is averaged over all radio mode phases the galaxy goes through and since it explicitly 
excludes AGN feedback which would blow away the gas, the Bondi-Hoyle accretion rate we use is a lower 
constraint. If we model AGN feedback, the derived accretion rates would be higher.

\subsubsection{Prolonged Gas Accretion Model}

The SMBH in our PGA model grows through a combination of black hole mergers and gas accretion, inspired 
by Micic et al. (2007).  To review the approach, we assumed that major galaxy mergers would funnel 
gas to the black hole in each of the progenitors and activate an Eddington-limited growth phase 
for a Salpeter time. The black hole mass in Micic et al. (2007) grew as: 
M$_{\rm BH}$(t) = M$_{\rm BH,0}$(t$_0$) exp($\Delta$t/t$_{\rm sal}$), where $\Delta$t = t - t$_0$, 
t$_{\rm sal} \equiv  \epsilon {\rm M}_{\rm BH} c^2 / [(1-\epsilon) {\rm L}]$, $\epsilon$ 
is the radiative efficiency, L is the luminosity, and c is the speed of light; in this picture 
the black hole mass would roughly double in 40 Myr (Hu et al. 2006). We distinguished two cases 
depending on the mass ratio of merging dark matter halos. The first is a more conservative 
criterion that allows black holes to accrete gas if the mass ratio of the host dark matter halo 
is less than 4:1 -- a major merger. The second case sets an upper constraint on the final black 
hole mass by allowing gas accretion as long as the merging dark matter halos have a mass ratio 
less than 10:1 -- on the cusp of what is considered a minor merger. Since our black holes merged 
promptly after the halos merged, the accretion timescale and efficiency for major mergers was 
the same regardless of the mass ratio or redshift.

In this paper, we continue to model the black hole growth as one of extended gas accretion excited 
by major mergers. At high redshift, this is likely a good assumption, though note that at low 
redshift when mergers are infrequent, secular evolution such as bar instabilities may dominate 
the gas (and therefore black hole) accretion. Integrated over the whole of a black hole lifetime, 
though, this major merger-driven gas accretion is likely to be the dominant source of gas inflow. 
Since the black hole growth is so strongly dependent on what fuel is driven to the center during 
galaxy mergers, it is important to characterize this merger-driven mass growth, including the critical 
gas physics that may inhibit or strengthen this nuclear supply. We are motivated by numerical 
simulations that include radiative gas cooling, star formation, and stellar feedback to study the 
starburst efficiency for unequal mass ratio galaxy mergers (Cox et al. 2008), which finds that the 
gas inflow depends strongly on the mass ratio of the galaxy (see also, e.g.,  Hernquist 1989, 
Mihos $\&$ Hernquist 1994). This study parametrizes the efficiency of nuclear star formation 
(i.e. gas supply and inflow), $\alpha$, as a function of galaxy mass ratio. In a broader sense, 
this study shows how much gas is available for either star formation or gas accretion onto the central 
black hole. These two processes compete for the same gas and the outcome (nuclear cluster versus SMBH) 
depends on the mass of the host spheroid (Nayakshin et al. 2009). The efficiency of gas inflow is 
described by:

\begin{equation}
{\alpha} = \alpha_{\rm slope} \Bigg({M_{\rm s} \over { M_{\rm p}}} - \alpha_0 \Bigg)^{0.5},
\end{equation}

\noindent where $\alpha_0$ defines the mass ratio below which there is no enhancement of nuclear 
star formation (i.e. gas inflow), and $\alpha_{\rm slope}$ is the fitted slope of the solid line
in Cox et al. (2008), Fig 15. Here, the gas accretion efficiency has a maximum of 0.56 for 1:1 halo 
mergers with $\alpha_{\rm slope}$=0.6, and falls to zero at $\alpha_0$. This parametrization is 
insensitive to the stellar feedback prescription. We use $\alpha$ to define how efficiently the 
merger funnels the galaxy's gas to the black hole accretion disk. In general, all feedback processes 
(stellar or black hole) can be contained in $\alpha_{\rm slope}$. Only small scale numerical 
simulations of gas accretion in AGNs, with fully implemented and resolved gas physics can relate 
$\alpha_{\rm slope}$ to the mass ratio of merging galaxies. In Cox et al. (2008), the fitted 
$\alpha_0$ parameter suggests gas inflow is sharply curtailed for mass ratios larger than 9. 
We set the cut off mass ratio to $M_{\rm p}/M_{\rm s} = 10$ in order to compare with our previous 
black hole growth prescription (Micic et al. 2007). We adopt $\alpha_{\rm slope} = 0.6$ as in 
Cox et al. (2008).

Now that we have implemented a realistic description of the merger time for each black hole within 
a halo (see next section), we allow them to grow for a physically-motivated accretion timescale. 
The accretion of gas onto both the incoming and central black hole starts when the two black holes 
are still widely separated, at the moment of the first pericenter passage, and continues until the 
black holes merge (c.f. Di Matteo et al. 2005, Colpi et al. 2007). This sets the accretion timescale, 
t$_{\rm acc}$, as follows: t$_{\rm acc}$ = t$_{\rm df}$(r=r$_{\rm vir}$) - t$_{\rm dyn}$(r=r$_{\rm vir}$), 
where t$_{\rm df}$(r=r$_{\rm vir}$) is the merger timescale including dynamical friction, and 
t$_{\rm dyn}$(r=r$_{\rm vir}$) is dynamical time at virial radius r$_{\rm vir}$, which marks the 
first pericenter pass of the black hole. By stopping the accretion as the black holes merge, we 
roughly model the effect of black hole feedback in stopping further accretion.

Putting these pieces together, the mass accreted by a black hole during 
t$_{\rm acc}$(r=r$_{\rm vir}$) is:

\begin{equation}
M_{\rm acc}=M_{\rm BH,0} (e^{\frac{\alpha {\rm t}_{\rm acc}}{{\rm t}_{\rm sal}}}-1),
\end{equation}

\noindent where M$_{\rm BH,0}$ is the initial black hole mass, $\alpha$ is the starburst efficiency 
(Cox et al. 2008), and t$_{\rm sal}$ is defined above. After t$_{\rm df}$, the incoming black hole 
merges with the SMBH at the center and a new SMBH is formed after having accreted gas for 
t$_{\rm acc}$. The accretion time and efficiency both implicitly encode the large-scale dynamics 
of the merger and the bulk gas accretion into the nuclear region, while t$_{\rm sal}$ describes 
the accretion disk physics. As before, we set the Salpeter accretion efficiency to 0.1 (Shakura 
$\&$ Syunyaev 1973).

\subsection{Dynamical Friction}

Dynamical friction allows massive black hole binaries to form at the center of a galaxy 
in two ways. First, dynamical friction expedites the merger of two dark matter halos and 
later the merger of the galaxies they host. In this manner, merging galaxies can efficiently 
shepherd massive black holes to the center of the new system, roughly to the inner kiloparsec 
(see Colpi et al. 2007 for a review). Second, dynamical friction from the gas in the disk 
carries black holes deeper toward the galactic center, where they form binary and eventually 
merge (e.g. Begelman, Blandford $\&$ Rees 1980, Escala et al. 2005, Kazantzidis et al. 2005, 
Dotti et al. 2007). We model both effects as follows:

The time for a satellite to sink to the center of a primary can be approximated using 
Chandrasekhar dynamical friction (Binney $\&$ Tremaine 1987):

\begin{equation}
t_{\rm chandra}=\frac{1.17}{{\ln}{\Lambda}}\frac{{\rm r_{\rm circ}^{2} 
v_{c} {\epsilon^{\alpha}}}}{{\rm GM_{\rm s}}},
\end{equation}

\noindent where $\ln\Lambda$ is the Coulomb logarithm, $\ln\Lambda$ $\approx$ 
$\ln {\rm (1 + M_{\rm p}/M_{\rm s})}$. To define the satellite orbit, we adopt 
values suggested by numerical investigations (Colpi et al. 1999) and used in 
previous semi-analytical work (Volonteri et al. 2003): the circularity
$\epsilon^{\alpha}$ = 0.8, and the circular velocity v$_{\rm c}$ is determined 
at r$_{\rm circ}$ = 0.6 r$_{\rm vir}$. For a 10$^{12}\Msun$ halo, r$_{\rm vir} \sim 300$ kpc.  

Assuming that each merging galaxy carries a massive black hole at its center, t$_{\rm fric}$ 
is the merging time for massive black holes when all other processes (3-body scattering, gas 
dynamical friction, gravitational radiation, etc.) involved in the formation and later shrinking 
of the black hole binary are efficient and fast. Due to tidal stripping and possible resonant 
interactions, simulations have shown that the Chandrasekhar formula underestimates the merger 
time, especially in the case of minor mergers (e.g. Holley-Bockelmann $\&$ Richstone 1999, 
Weinberg 1989). If this is true, then semi-analytic studies of black hole merger rates using a 
Chandrasekhar formalism for the merger time will overestimate the true number of black hole mergers.

In an effort to better parametrize dynamical friction, Boylan-Kolchin et al. 2008 used N-body 
simulations to study dark matter halo merging timescales, and confirmed that the Chandrasekhar 
formalism does underestimate the merger time, by a factor of $\approx$ 1.7 for 
M$_{\rm p}$/M$_{\rm s}$ $\approx$ 10 and a factor of $\sim$ 3.3 for M$_{\rm p}$/M$_{\rm s}$ $\approx$ 100. 
They propose a fitting formula that accurately predicts the timescale for a satellite to sink 
from the virial radius to the host halo center: 

 \begin{equation}
   \frac{\tdf}{\tau_{\rm dyn}} = A \, {(\mhost/\msat)^b \over \ln(1+\mhost/\msat)}
  \exp\left[c \, {j \over j_c(E)} \right] \, \left[{r_c(E) \over \rvir} \right]^d,
\end{equation}
where b = 1.3, c = 1.9, d = 1, A = 0.216, circularity 
j / j$_c$(E) = 0.5, r$_c$(E) / r$_{\rm vir}$ = 0.65, as defined in Boylan-Kolchin 
et al. 2008. The dynamical time, t$_{\rm dyn}$, is given at virial radius as:

\begin{equation}
   \tau_{\rm dyn}\equiv \frac{\rvir}{V_c(\rvir)}
           =\left( \frac{\rvir^3}{G\mhost} \right)^{1/2} \,,
\end{equation}
where V$_{\rm c}$(r$_{\rm vir}$) = (GM$_{\rm vir}$/R)$^{1/2}$.

\begin{figure}
\vspace{5mm}
\begin{center}
\includegraphics [height=2.5in,width=3.in,angle=0]{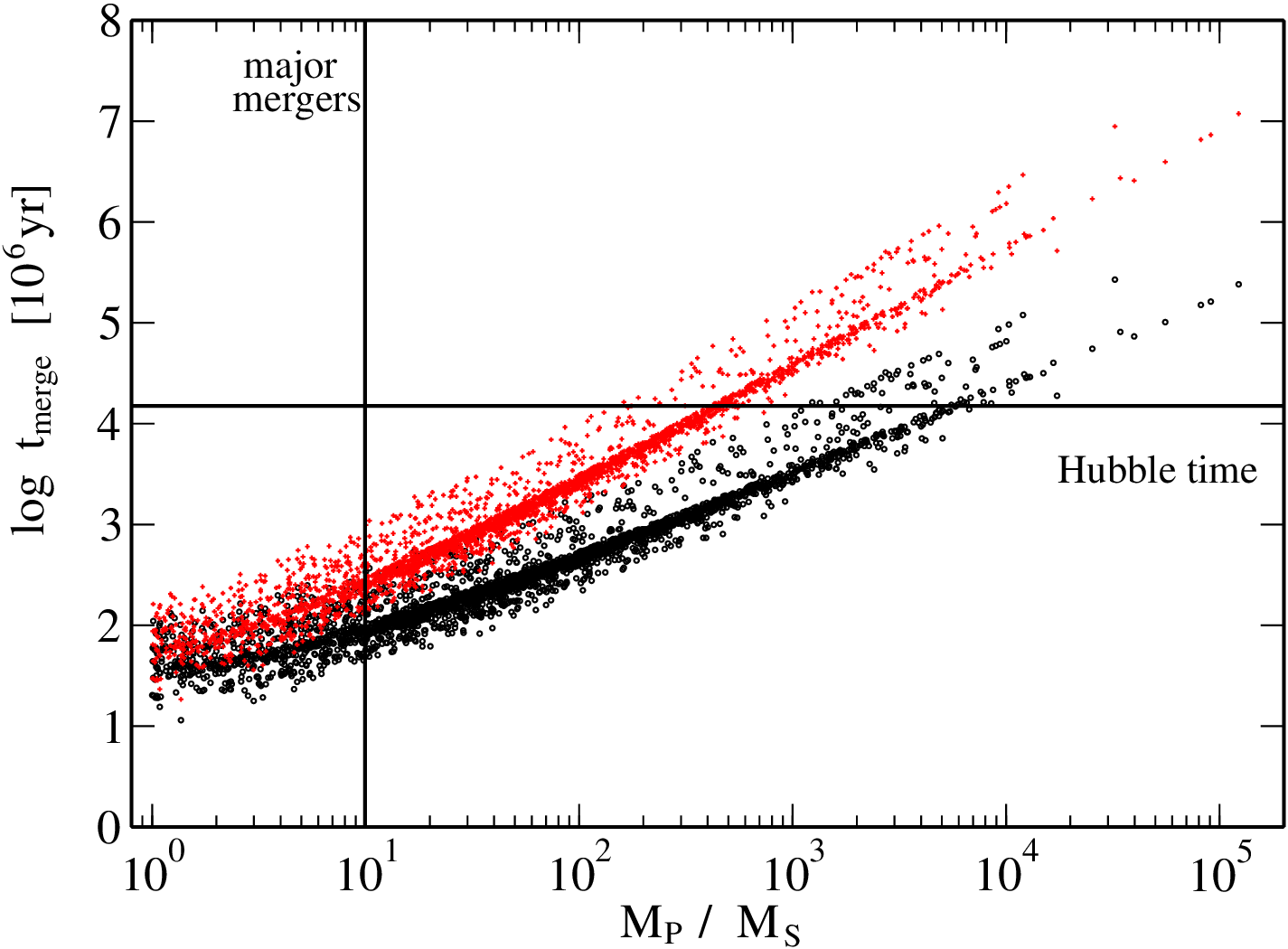}
\caption[Fig 1.]{Time for each satellite to reach the primary halo center as a 
function of the mass ratio of merging halos. Black circles represent the Chandrasekhar 
dynamical friction time and red pluses represent the merging time calculated from a 
simulation-based numerical fit (Boylan-Kolchin 2008). Both timescales are compared to 
the Hubble time and mass ratios are compared to major mergers. Merging halos above the 
horizontal black line will not finish their merger and are removed from the merger tree, 
but their positions are updated within the primary halo.}
\end{center}
\end{figure}

Figure 1 shows the dark matter halo merger timescale for each pair of merging halos for both 
dynamical friction estimates. We have calculated the dark matter halo merger rate with 
both the Chandrasekhar dynamical friction formula and the Boylan-Kolchin numerical fit 
(Holley-Bockelmann et al. 2010). Unless otherwise stated, we adopt the 
Boylan-Kolchin timescale for all calculations.

It is expected that in gas-rich galaxies, dynamical friction from the gas would bring two black 
holes close enough to form a binary whose orbit would shrink efficiently, passing quickly from a 
binary in the 3-body scattering phase to one emitting significant gravitational radiation (Escala 
et al. 2005, Kazantzidis et al. 2005, Dotti et al. 2007). Numerical simulations indicate that two 
black holes will sink from $\sim$ 1 kpc to form a binary with a separation of less than a parsec 
in $\sim$ 10 Myrs. We incorporate this physics by calculating the dynamical friction timescale 
from the virial radius to the inner kpc, and then assume that the two black holes merge 10 Myr 
afterward. In practice, the dynamical friction timescale from equations 1 and 2 from the inner kpc 
to the bound binary stage is often of order 10 Myr; the power of this gas-rich assumption lies in 
that it entirely circumvents the so-called 'final-parsec' problem thought to exist for low mass 
ratio mergers of $10^{6.5} \Msun$ $\leq$ M$_{\rm BH}$ $\leq 10^8 \Msun$ within static, spherical, 
gas-poor galaxy models (e.g. Milosavljevic $\&$ Merritt 2003). We explicitly assume that the black 
holes in our simulation do not stall at the final parsec before merger, and instead are ushered 
efficiently into the gravitational radiation stage where they coalesce; this assumption may hold 
even in the case of gas-poor galaxy models as long as the model is not spherical or in 
equilibrium (e.g. Holley-Bockelmann $\&$ Sigurdsson 2006, Berczik et al. 2006, R. Spurzem, private 
communication). Note that this implies that if we assume that each halo initially carries a black 
hole at its center, and that each host galaxy is gas-rich, figure 1 also estimates the time for a 
black hole to sink from the virial radius to the center of the host galaxy, become a bound black 
hole binary, and inspiral due to gravitational radiation.

In our initial work (Micic et al. 2007), mergers of dark matter halos trigger the immediate merger 
of the black holes they are hosting. In this paper, subsequent mergers of the central black holes 
are delayed to account for dynamical friction of the halos and the black holes within the galaxy. 
Black holes will not merge if their merger time is larger than a Hubble time, and in that case, 
we advance the black hole position within the primary halo at each timestep. Knowing the dynamical 
friction timescale for each merger, we postpone the black hole mergers accordingly. For the final 
kpc, we assume that ambient gas and/or non-sphericity will cause two black holes to coalesce within 
10 Myrs.

\subsection{Gravitational Recoil}

Binary black holes strongly radiate linear momentum in the form of gravitational waves during the 
plunge phase of the inspiral -- resulting in a ``kick'' to the new black hole. This, in itself, 
has long been predicted as a consequence of an asymmetry in the binary orbit or spin configuration. 
Previous kick velocity estimates, though, were either highly uncertain or suggested that the 
resulting gravitational  wave recoil velocity was relatively small, astrophysically speaking. Now, 
recent results indicate the recoil can drive a gravitational wave kick velocity as fast as 
$\sim$4000$\kms$ (Herrmann et al. 2007, Baker et al. 2007, Gonzalez et al. 2007a, 2007b, Koppitz et al. 2007, 
Campanelli et al. 2007, Schnittman $\&$ Buonanno 2007, Lehner $\&$ Moreschi 2007, McWilliams 2008, 
Lousto $\&$ Zlochower 2009, Sperhake 2009). In reality, much smaller values than this 
maximum may be expected in gas-rich galaxies due to the alignment of the orbital angular momentum
and the spins of both black holes (Bogdanovic et al. 2007). \footnote{However, recent observations 
(Komossa et al. 2008) may show evidence for large kick velocities $\sim$ 2500 $\kms$.} Recent studies 
also hint at a potential purely general relativistic spin alignment mechanism (Kesden et al. 2010). 
However, even typical kick velocities ($\sim 200~\kms$) are interestingly large when compared to the 
escape velocity of most astronomical systems -- low mass galaxies, as an example, have an escape 
velocity of $\sim 200~\kms$ (e.g. Holley-Bockelmann et al. 2007). The effect of large kicks combined 
with a low escape velocity from the centers of small dark matter halos at high redshift may play a 
major role in suppressing the growth of black hole seeds into SMBHs. Even the most massive dark matter 
halo at z$\geq$11 can not retain a black hole that receives $\geq$ 150 ${\rm km \,s^{-1}}$ kick 
(Merritt et al. 2004, Micic et al. 2006). We incorporate the effect of recoil velocity on the growth 
of the SMBH by assigning a kick to each black hole merger in our merger tree. We follow the approach 
adopted by Holley-Bockelmann et al. (2007), which uses the parametrized fit of Campanelli et al. 2007 
to generalize the recoil velocity as a function of the mass ratio of the merging black holes, each 
individual black hole's spin amplitude, and the alignment to the orbital angular momentum. We assume 
that as black holes form a hard binary, their orbits will be highly circular and the eccentricity 
is close to zero:

\begin{equation}
v_{\mathrm{kick}} = [(v_{m}+v_{\perp} cos{\xi})^2+(v_{\perp} \sin{\xi})^2+(v_{\parallel})^2]^{1/2},  
\end{equation}
where
\begin{equation}
v_{m} = A \frac{q^2 \left(1-q\right)}{\left(1+q\right)^5} \left[ 1+
  B \frac{q}{\left(1+q\right)^2}\right], 
\end{equation}
\begin{equation}
v_{\perp} = H \frac{q^2}{\left(1+q\right)^5} \left( \alpha_2^\parallel
  - q \alpha_1^\parallel\right), 
\end{equation}
and
\begin{equation}
v_{\parallel} = K \cos\left(\Theta-\Theta_0\right) \frac{q^2}{\left(1+q\right)^5}
\left( \alpha_2^\perp - q \alpha_1^\perp \right).
\end{equation}
Here, the fitting constants are A$=1.2 \times 10^{4} \kms$, B$=-0.93$, H$=(7.3 \pm 0.3) \times 10^{3} \kms$, 
and K$\cos\left(\Theta-\Theta_0\right) =(6.0 \pm 0.1) \times 10^{4} \; $, while the subscripts $1$ and $2$ 
refer to the first and second BH respectively; $\perp$ and $\parallel$ stands for perpendicular and parallel 
to the orbital angular momentum; the mass ratio q$\equiv {\rm M}_2/{\rm M}_1$; the reduced spin parameter 
$\alpha_i \equiv {\rm S}_i/{\rm M}_i^2$, where S$_i$ is the spin angular momentum of BH $i$. The orientation 
of the merger is specified by: $\Theta$, the angle between the ``in-plane'' component of 
$\delta^i \equiv ({\rm M}_1 + {\rm M}_2) \left({\rm S}_2^i/{\rm M}_2  - {\rm S}_1^i/{\rm M}_1\right)$ and the 
infall direction at merger; $\Theta_0$, the angle between $\delta^i$ and the initial direction of motion; and 
$\xi$, the angle between the unequal mass and spin contribution to the recoil in the orbital plane. 

We assume that the orbit is circular, and we take the mass ratio of merging black holes directly from our 
merger tree. We have the following free parameters: the spin amplitude and orientation of each black hole, 
and the orientation of the merger. We explore two spin distributions. The K1 model chooses the spin parameters 
from a uniform distribution,  while the K2 model assumes the black hole spins are aligned with the orbital 
angular momentum (Bogdanovic et al. 2007). Figure 2 shows the distribution of kick velocities for both models. 
We apply both kick velocity distributions to the mergers in our merger tree using 1000 realizations each. 
This yields 2,000 Milky Way merger trees for the PGA model and 2,000 for the FPL model.

\begin{figure}
\vspace{5mm}
\begin{center}
\includegraphics [height=2.5in,width=3.in,angle=0]{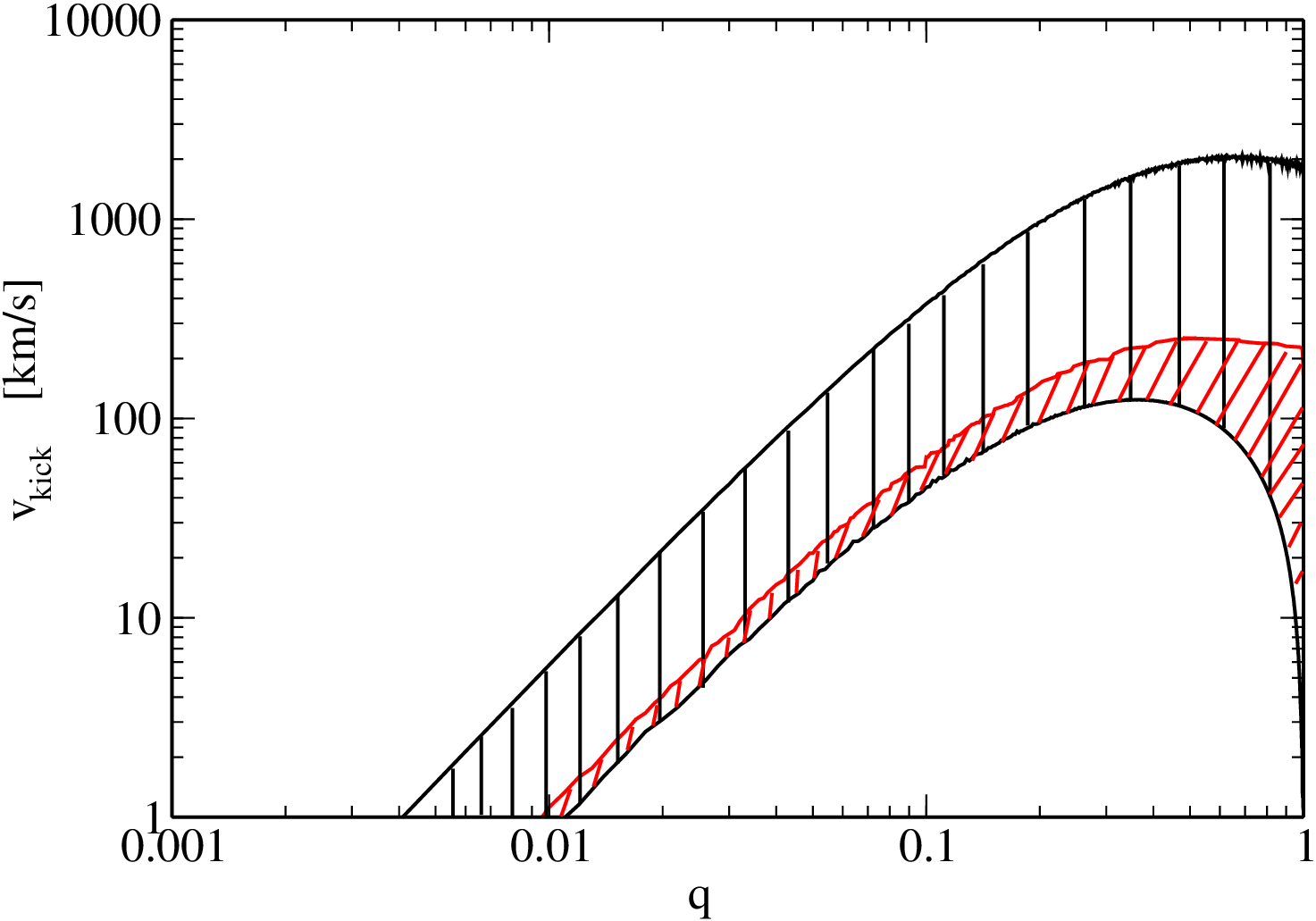}
\caption[Fig 2.]{Distribution of gravitational recoil as a function of the mass ratio 
of merging black holes on circular orbits. The black lines represent random spin 
orientations and amplitudes -- the K1 distribution. In red is the kick range when the 
spin orientation of the binary black holes aligns with the orbital momentum of the 
binary  -- the K2 distribution. If  the kick is larger than the escape velocity, the 
black hole is ejected from the halo.} 
\end{center}
\end{figure}

For each black hole merger, we calculate the kick velocity and compare it to the escape velocity 
of the host halo at the time of merger. If the kick is larger than the escape velocity, the 
resulting black hole is removed from the center of the host halo and from the merger tree. 
As the result, the host halo will not have a central black hole for a period of time. Subsequently, 
when the next black hole sinks to the center, it will simply take place of the central black hole. 
Depending on how often the central black hole is ejected, the final SMBH mass may be substantially 
smaller. The sequence and number of kicks in one merger tree realization depends on our kick distribution. 
Over many realizations, we can produce a probability function for the final black hole mass for each model.

\section{RESULTS}

Black hole mergers postponed for longer than a Hubble time will not occur, which reduces the 
merger rate over all redshifts. Those mergers that are postponed but have a merging timescale less 
than a Hubble time will merge at lower redshifts. This is made explicit in figure 3, which plots 
the average change in redshift for a merger as a function of redshift if one includes dynamical 
friction. Black hole mergers at redshift 10 are, on average, pushed to redshift 7, for example. 
This results in an increase in the black hole merger rate at low redshifts. Since each merger 
occurs at a lower redshift, it will certainly be a louder gravitational wave source, and any 
associated electromagnetic signature will be brighter, as well (Holley-Bockelmann et al. 2010).

\begin{figure}
\vspace{5mm}
\begin{center}
\includegraphics [width=2.7in,angle=0]{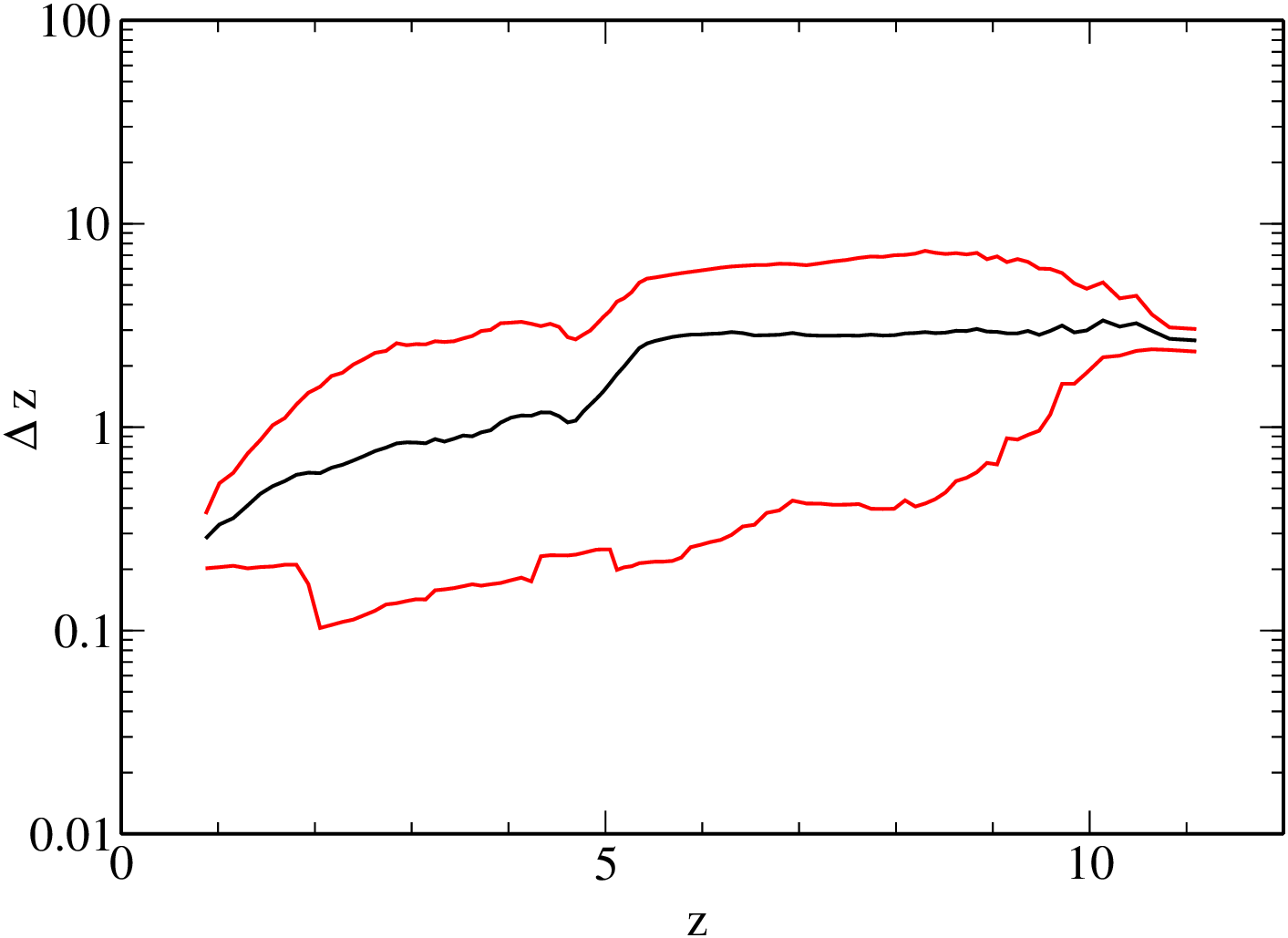}
\caption[Fig 3.]{Thick black - the average change in the merger redshift when dynamical 
friction is applied as a function of merger redshift (excluding dynamical 
friction). Red lines represent the minimum and maximum shift in the black hole merger 
redshift. Dynamical friction postpones black hole mergers toward lower redshifts, making 
them louder LISA sources and increasing the local rate as well.}
\end{center}
\end{figure}

Interestingly, including dynamical friction from gas at distances smaller than 1 kpc does not 
make difference in our merger time scales. Those black holes that reach 1 kpc from the galactic 
center will reach a parsec even without the gas -- gas is simply invoked to ensure 
that the black holes pass through the 3-body scattering stage efficiently. For the rest of this 
paper, we adopt the Boylan-Kolchin dynamical friction fit as a realistic treatment of the black 
hole merger timescale.

\subsection{Realizations of Supermassive Black Hole Growth}

Figure 4 shows the merger rates in our volume for those dark matter halos that merge in less 
than a  Hubble time within the mass range 10$^7\Msun$ $\leq$ M$_{\rm DMH}$ $\leq$ 10$^{13}\Msun$, 
for various mass ratios and combined masses of merging dark matter halos. In our volume, all 
the mergers that occur at redshifts z $\leq$ 1.5 will not finish in less than a Hubble time, as 
these mergers are all high mass ratio. We remind the reader, though, that these halo merger 
rates are calculated for VL-2 40 Mpc$^3$ simulation box. This model, by design, has no Galaxy Clusters, 
or even rich local groups, and it is for sparse local groups. The VL2 host halo was specifically 
chosen ``not'' to have undergone a major merger more recently than about z$\sim$1. 
Almost all the low redshift mergers for this mass range SMBH are in richer, denser environments, 
which are just not covered in this model. A better global rate for these low mass halos will be 
achieved with more of these volumes.

\begin{figure}
\vspace{5mm}
\begin{center}
\includegraphics [height=4.in,width=2.7in,angle=0]{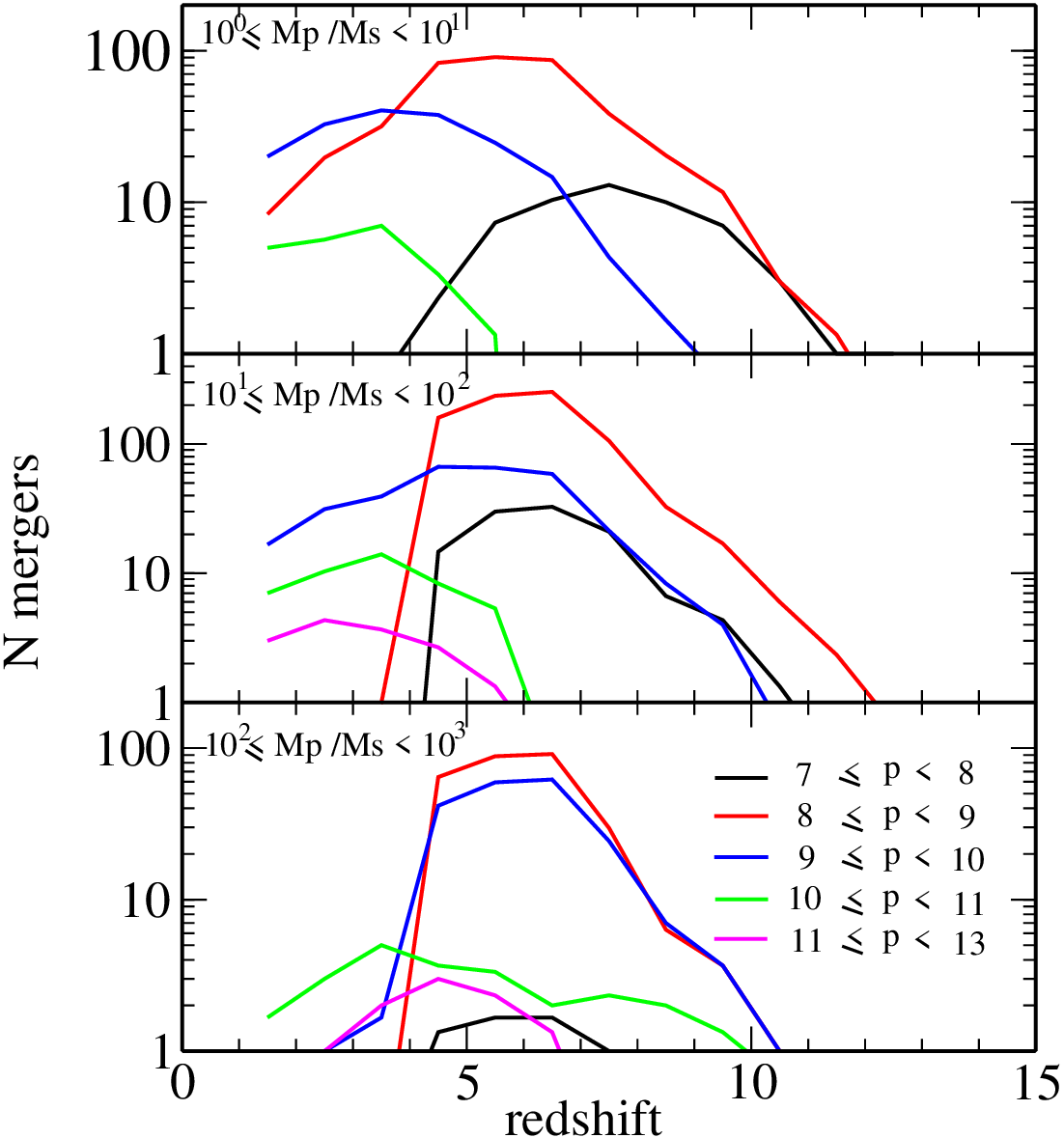}
\caption[Fig 4.]{Dark matter halo merger rates as a function of redshift for all halos
that finish merging by redshift zero. Three panels show various ranges for the halo 
mass ratio and the total combined halo mass. M$_P$ is the primary halo mass, 
M$_S$ is the satellite mass, and the total mass is defined by p = log (M$_P$ + M$_S$). 
The panel with M$_P$/M$_S$ $\leq$ 10 shows merger rates for major mergers that activate 
gas accretion onto their black holes.}
\end{center}
\end{figure}

Since in PGA the merger timescale is larger for higher mass ratio halos, the black hole gas accretion 
time scale is longer as well. Figure 5 shows this time scale for black holes 
hosted by dark matter halos that merge with mass ratio M$_{\rm p}$/M$_{\rm s}$ and combined mass 
p=log(M$_{\rm p}$+M$_{\rm s}$). Notice that gas accretion is activated only for a small range of 
halo mass ratios M$_{\rm p}$/M$_{\rm s}$ $\leq$ 10 (the dashed line in figure 5). For these 
mergers, the black holes do not accrete for longer than $\sim$ 1000 Myr, and the accretion 
efficiency is strongly damped as the halo mass ratio tends toward 10:1 -- in no case does the system 
reach the Eddington limit.

\begin{figure*}
\vspace{10mm}
 \epsfig{figure=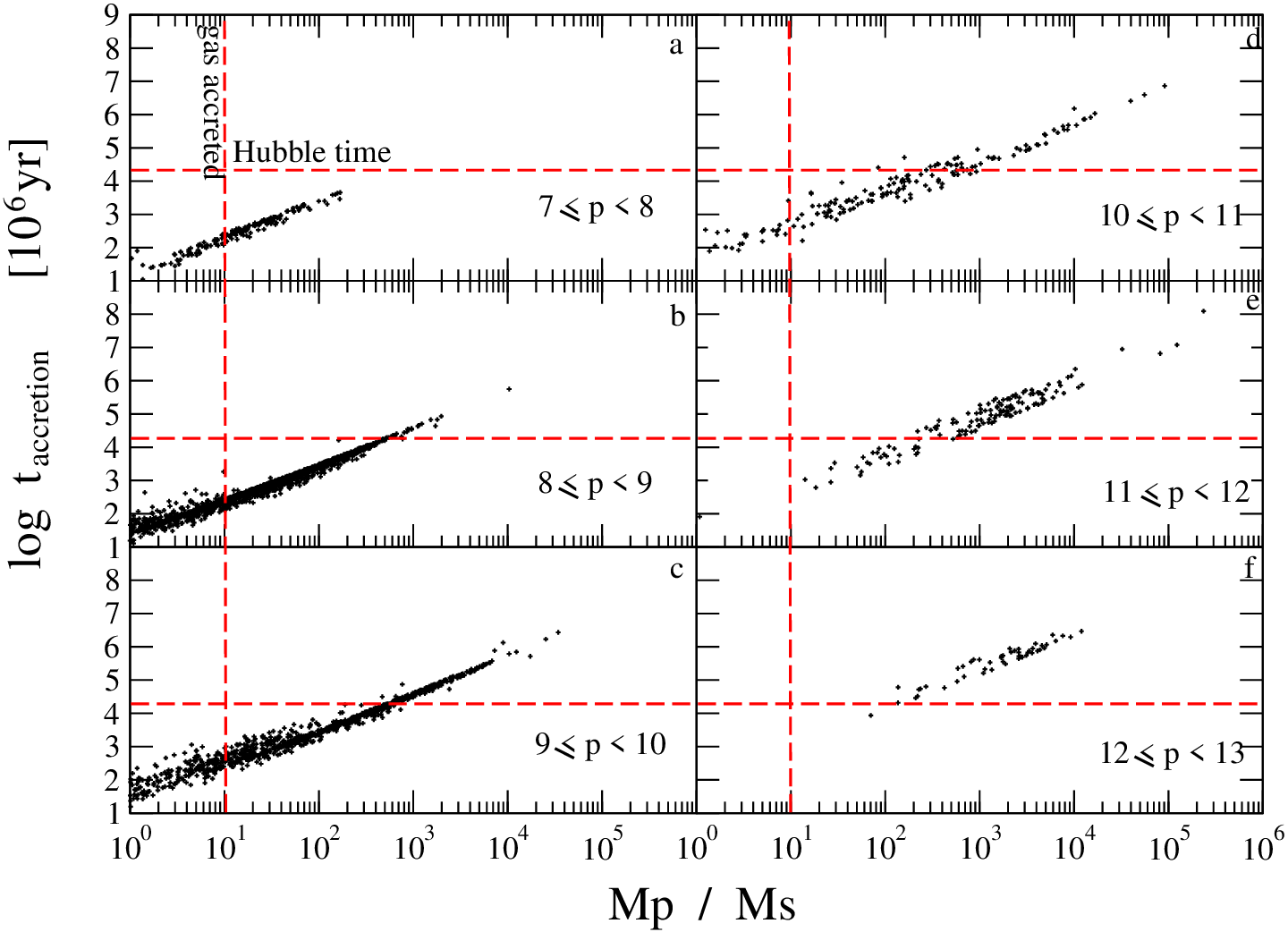, height=0.990\textwidth, width=0.990\textwidth}
 \vspace{5mm}
 \caption{Gas accretion time, $t_{\rm acc}$, for a central black hole in a 
satellite halo of mass M$_{\rm sat}$ that is merging with a primary halo of mass 
M$_{\rm prim}$ in the PGA model. The total mass is denoted by p = log (M$_{\rm s}$ +M$_{\rm prim}$), and 
t$_{\rm acc}$ = t$_{\rm merger}$ - t$_{\rm dyn}$ where t$_{\rm merger}$ is the merger time 
scale, and t$_{\rm dyn}$ is the dynamical time,  making accretion begin at the first 
pericenter pass. Halos that merge with mass ratios less than 10 (vertical line) will 
allow gas accretion onto the black holes. Only satellite halos with mass $\leq$ 10$^{11}\Msun$
will have an accreting central black hole.}

\end{figure*}

Figure 5 offers valuable information about the FPL model as well. Panels e and f  show 
that there are no mergers with mass ratios 10:1 or less for halos above 10$^{11}\Msun$. In fact, 
the halo that VL-2 marks as the Milky Way halo does not go through any major mergers at low 
redshift. In the context of FPL,  the quasar mode of rapid gas accretion occurs in Milky Way 
progenitors at high redshifts while the radio mode dominates most of the Sgr A* growth at low 
redshifts. We should also point out that VL-2 merger history of Milky Way is just one of the 
possible merger histories, and perhaps not the true one.

\subsubsection{The Initial Mass Function of Black Hole Seeds.}

Galaxies in the mass range of the VL-2 simulation have values for escape velocities comparable to the 
amplitudes of gravitational wave recoil for black hole mergers with mass ratios q$\geq$0.1, even in 
the case of the lower K2 kick distribution. If we are to seed these galaxies with their first black holes, 
the choice of the black hole mass function (BHMF) will determine the mass ratio of merging black holes. 
We examine the influence of the choice of BHMF on VL2 merger trees in 1,000 kick realizations for BHMFs 
with constant values of 100; 1000; and 5000 $\Msun$ seeds; and also for random values in the mass ranges 
10 - 200 $\Msun$; 10 - 500 $\Msun$; and 10 - 1,000 $\Msun$. We find that having a constant BHMF or one 
in a narrow mass interval leads to close-to-equal mass ratio black hole mergers. These mergers have the 
highest gravitational wave recoil and ejects of black holes in high redshift dwarf galaxies, which
suppresses SMBH growth. Due to the uncertainty in the formation channel for seed black holes, as well as 
the lack of theoretical constraints on primordial gas fragmentation during PopIII star formation 
(Glover et al. 2010), 10 - 1,000 $\Msun$ is the least biased of our BHMFs. We adopt this flat BHMF 
hereafter for the seeds in our merger trees in both the FPL and PGA models. Note that the seed mass 
in the FPL model is much less relevant than in the PGA model, because the black hole can not grow more 
massive than what is dictated by the black hole fundamental plane, regardless of the initial seed mass.

\begin{figure}
\vspace{5mm}
\begin{center}
\includegraphics [height=3.9in,width=2.4in,angle=0]{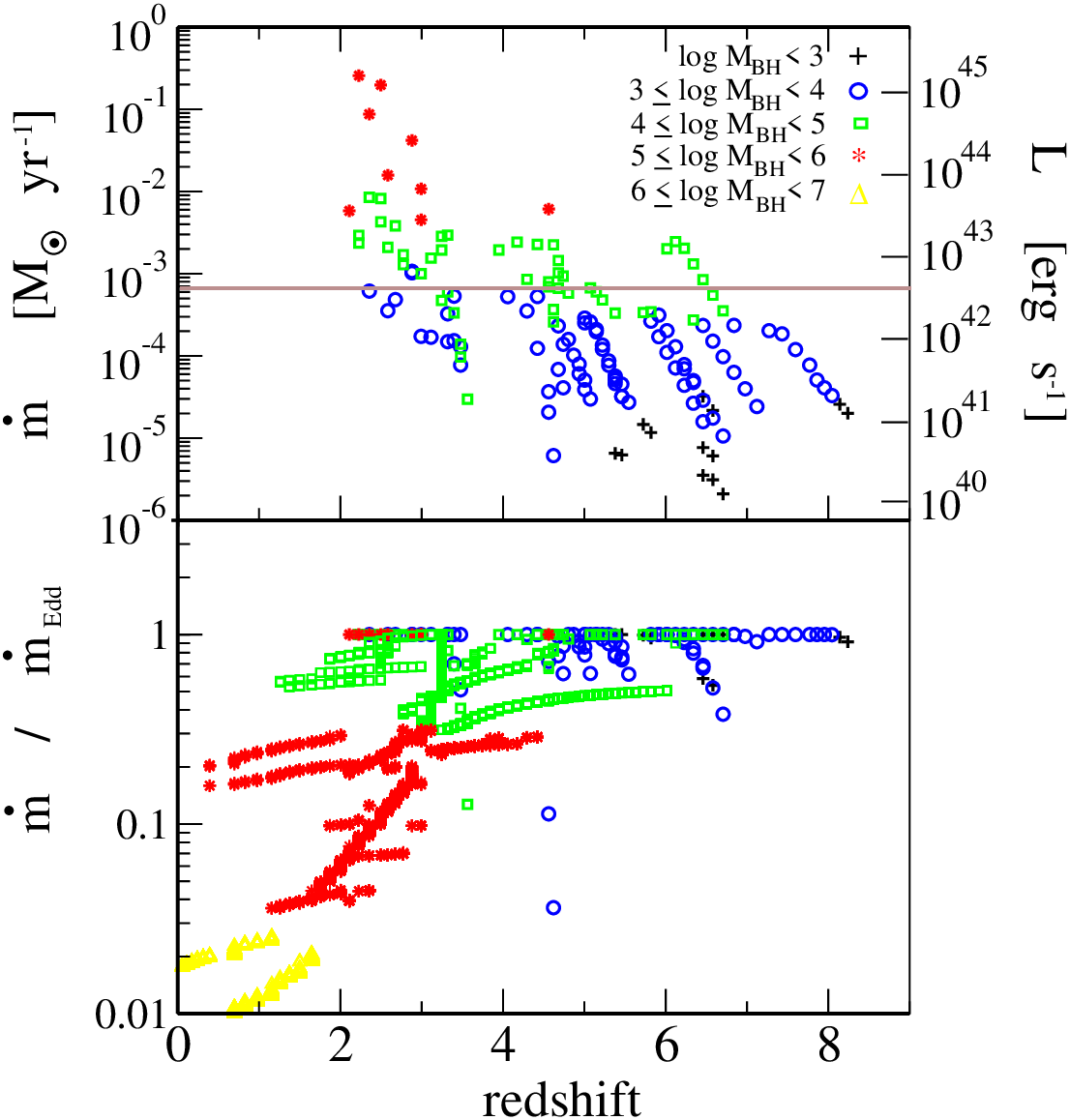}
\caption[Fig 6.]{Accretion rates and luminosities during the quasar and radio modes 
for the FPL model (recoil excluded). The upper panel shows gas accretion rates and luminosities for all 
black holes during a random short time interval ($\delta$t = t$_f$ - t$_i$). Points 
represent the quasar mode, while the brown horizontal line represents the radio 
mode. The black hole mass in the quasar mode increases at lower redshifts, with the final 
accretion episode between z=3 and z=2. Bondi - Hoyle accretion is the dominant growth 
mechanism at redshift z $\leq$ 2. The bottom panel shows the gas accretion rate at short 
time intervals compared to the Eddington rate (quasar and radio modes combined). Most 
black holes accrete at the Eddington rate by design. }
\end{center}
\end{figure}

\subsubsection{Massive Black Hole Accretion Properties}

With the adopted flat BHMF, the only constraints in the FPL model are the time-average 
Bondi-Hoyle accretion rate during the radio mode and the strength of gravitational wave 
recoil. We start by running Milky Way merger trees for various values of the accretion 
rate and recoil distribution, looking for the values which, in combination with the 
accretion during quasar mode, produce a $\sim$ 4.2 $\times 10^6\Msun$ black hole at z=0 in 
the majority of merger tree realizations. This condition is satisfied for the merger trees 
with $\dot m$$_{\rm bondi}$ = 7 $\times10^{-4}\Msun$yr$^{-1}$ (brown horizontal line in figure 6 up), 
consistent with the observations 
of accretion rates in low mass AGNs in the local Universe. It has been suggested that the 
Milky Way has evolved from a similar low mass AGN, which explains the fact that the observed 
value for accretion in Sgr A* today is an order of magnitude smaller than this best fit 
accretion rate. Outside of the Sgr A* analogue, there are also 35 field dwarf galaxies at z=0 with 
central black holes (see section 4.1). Figure 6 shows black hole accretion properties during 
the quasar and radio mode for the merger tree that excludes recoil, which corresponds 
to the maximum accretion. In this model, Milky Way progenitors go through their last 
quasar mode between z=2 and z=3 (upper panel, Fig 6), after which Bondi-Hoyle accretion 
becomes the dominant mechanism of Sgr A* growth (horizontal line in upper panel, Fig 6). 
In the bottom panel of figure 6, the same accretion rate is compared to the Eddington rate. 
Most of the accretion occurs at the Eddington rate -- this is expected since the assumption 
is built in that the black holes reach the Eddington rate very quickly. In many cases, however, 
accretion is not activated at all because the host galaxy is small and the spheroid mass
prevents the growth of central black hole to match the black hole fundamental plane.

\begin{figure*}
\vspace{10mm}
 \epsfig{figure=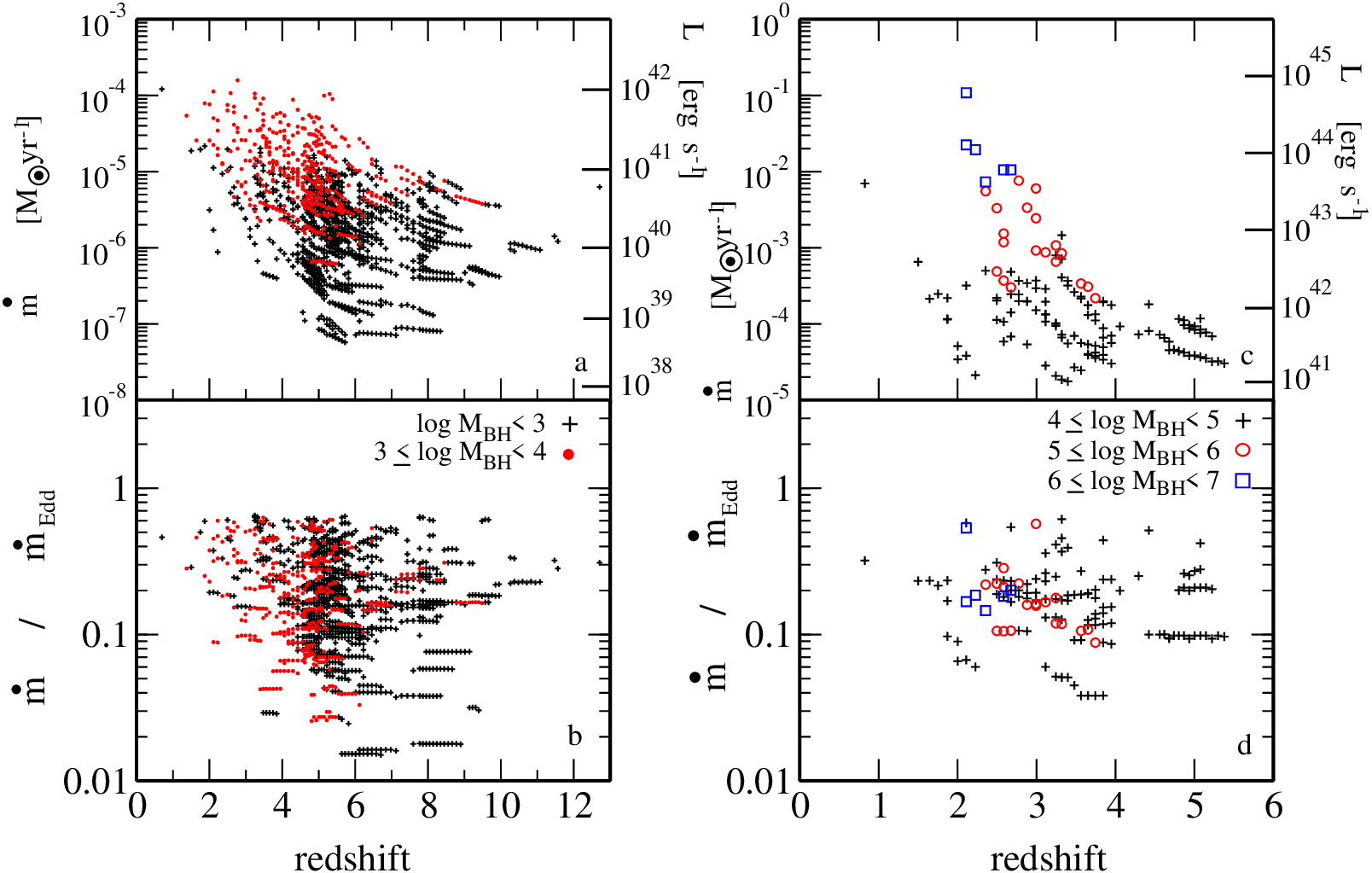, height=0.740\textwidth, width=0.990\textwidth}
 \vspace{5mm}
 \caption{Accretion in the PGA model. Left panels show black holes with mass less 
than $10^4\Msun$, and right panels show more massive black holes. Upper panels: Gas accretion 
rates and luminosities for all accreting black holes during a random short time interval 
($\delta$t = t$_f$ - t$_i$). Bottom panels: Gas accretion rates at short time intervals as a 
fraction of the Eddington rate for all accreting black holes. Unlike the FPL model, here most 
black holes accrete well below the Eddington rate. There are more ($\leq$ 10$^4\Msun$) accreting 
black holes in the PGA model because their growth is not limited by the black hole fundamental 
plane. }

\end{figure*}

By excluding kicks, figure 7 shows the upper constraint on black hole accretion properties 
for all black holes in the PGA merger tree. Recall that the BHMF is flat and the accretion 
efficiency is set by simulations. With the choice of BHMF and gas inflow efficiency fixed 
in this model, kicks are the only free parameter. Black holes in the PGA model can grow 
beyond the limit specified by the black hole fundamental plane. Instead of invoking AGN 
feedback in the form of an upper limit on the black hole mass, we incorporate it into the 
gas accretion efficiency. These feedback mechanisms suppress the accretion efficiency, 
leading to accretion rates that are 0.01 - 0.8  of Eddington (bottom panel, Fig 7). these 
lower accretion efficiencies also result in lower luminosities in the PGA model. While for 
$\sim$10$^6\Msun$ black holes, the  maximum luminosities is $\sim$ 10$^{45}\Lsun$ in the FPL 
model (upper panel, Fig 6), these black holes have an order of magnitude lower luminosity in 
the PGA model (upper panel, Fig 7). The maximum accretion 
rate of the PGA black holes at the redshift of their last accretion is $\sim$ 0.01 - 0.1 
$\Msun$/year or 0.8 $\dot m$$_{\rm Edd}$. Thereafter, the gas accretion is extremely damped. 
Although these SMBHs are far too low a mass to be considered a high luminosity AGNs analogue, 
it may be appropriate to link these SMBHs with low luminosity AGNs (Ho 2008). Our approach is 
in broad agreement with the observations of AGN lifetimes (Hopkins $\&$ Hernquist 2009), which 
show that most local black holes must have gained their mass in no more than a couple of accretion 
episodes. Again, we point out that Hopkins $\&$ Hernquist 2009 discuss their results for more 
massive black holes ($\geq$10$^{7.5}\Msun$) than those in our study.

\subsubsection{Black Hole Growth with Gravitational Wave Recoil}

We include gravitational wave recoil from the uniform K1 and lower recoil K2 distributions and reproduce 1,000 
merger trees for each kick distribution in both the FPL and PGA models. This results in 4,000 merger 
trees presented in figures 8, 9, 10 ,and 11.

\begin{figure}
\vspace{5mm}
\begin{center}
\includegraphics [width=3.in,angle=0]{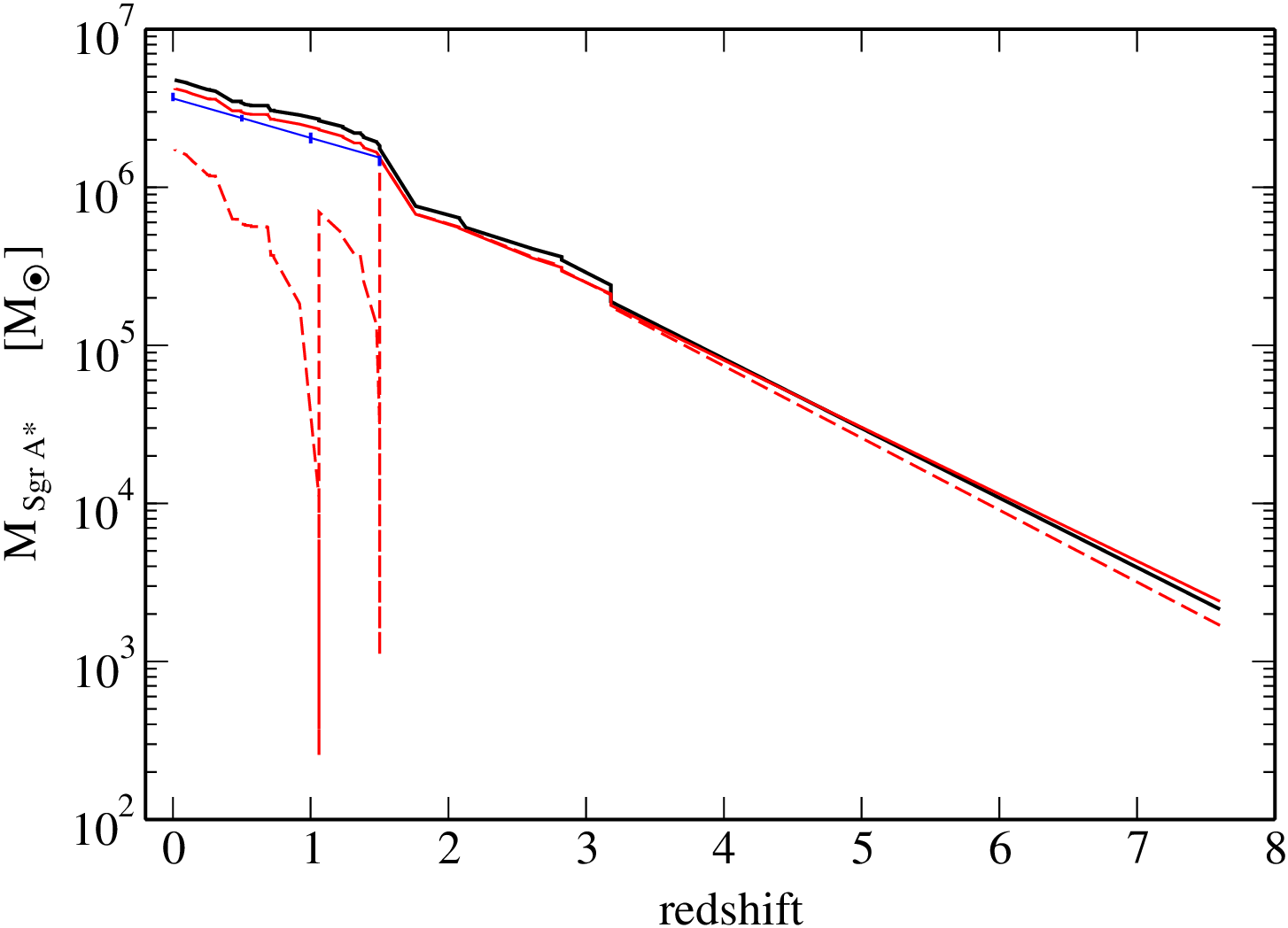}
\caption[Fig 8.]{The evolution of Sgr A* in the FPL model for 1000 K1 kick realizations. The black solid
line shows the black hole growth when kicks are excluded. Solid and dashed red lines show 
the black hole growth that produces the largest (4.3$\times$10$^6\Msun$) and smallest (2.0$\times$10$^6\Msun$) 
final black hole. The most common black hole mass is 3.9$\times$10$^6\Msun$ (blue). Also in blue: 
the 1-$\sigma$  spread around the most common value for redshifts z=1.5; 1.0; 0.5; and 0.0, further 
showing that the most common mass is near the mass of Sgr A*. }

\end{center}
\end{figure}

Figure 8 shows results for the K1 distribution in the FPL model. When kicks are excluded, the final 
black hole mass is $5 \times 10^6 {\rm M}_{\odot}$. The largest and smallest black holes at the center 
of the Milky Way analog are 4.3$\times$10$^6\Msun$ and 2.0$\times$10$^6\Msun$, respectively, while the 
most common black hole mass is 3.9$\times$10$^6\Msun$. Figure 8 also shows the 1-$\sigma$  spread around 
the most common value for redshifts z=1.5; 1.0; 0.5; and 0.0, further showing that the typical mass 
remains near the mass of Sgr A*. In the realization that favors the smallest SMBH mass, the kicks are 
so large that the black hole is ejected from the center of the Milky Way in two occasions. The reason 
the FPL model produces a final SMBH mass that so closely matches Sgr A* is quite trivial: it is built 
into the model that any major merger will grow the SMBH up to the mass dictated by the black hole fundamental 
plane. Even when the central black hole in the Milky Way analog is only $\sim 100 {\rm M}_{\odot}$ at z=1.1, 
there is plenty of time to grow the black hole back to the supermassive range.

\begin{figure}
\vspace{5mm}
\begin{center}
\includegraphics [width=3.in,angle=0]{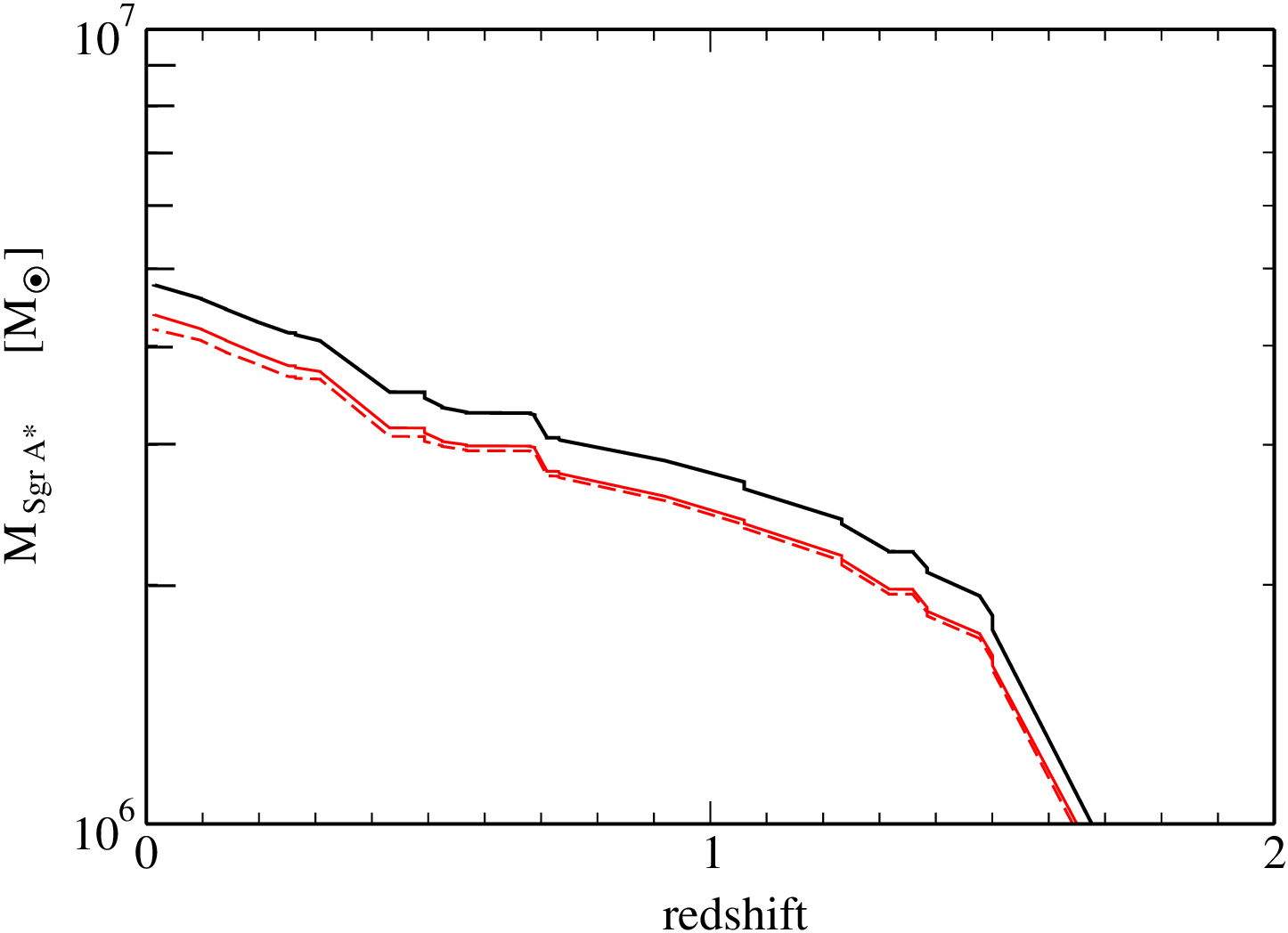}
\caption[Fig 9.]{Sgr A* evolution in the FPL model for 1000 K2 kick realizations (lower recoil). 
Black : the z $\leq$ 2 part of the merger tree when kicks are excluded. Solid red:  
kick realization favoring the largest black hole at the center of Milky Way analog. 
Dashed red: the kick realization favoring the smallest final black hole. Both the smallest 
and largest values are close to the observed Sgr A* mass.}      

\end{center}
\end{figure}

This also acts to shift the typical black hole mass toward the merger tree realization with 
largest final black hole mass. In the case of lower amplitude K2 distribution (with black hole 
spin axes aligned) this effect is even more pronounced (figure 9). In figure 9, we show the 
z $\leq$ 2 part of the merger tree for the case that excludes kicks, as well as the
extrema in final black hole masses, all of which are close to the observed Sgr A* mass.

\begin{figure}
\vspace{5mm}
\begin{center}
\includegraphics [width=3.in,angle=0]{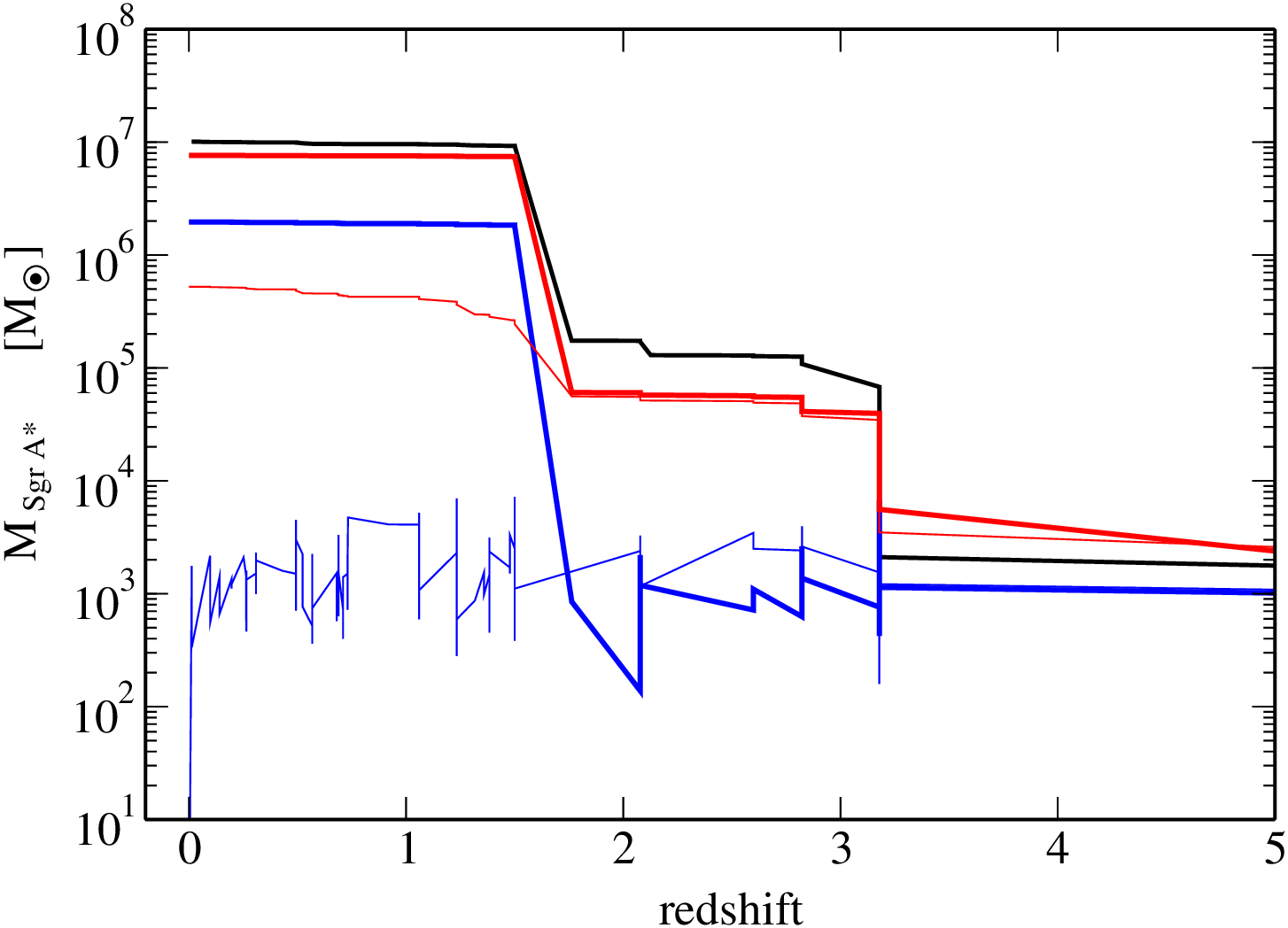}
\caption[Fig 10.]{Sgr A* evolution in the PGA model for 1000 K1 and K2 (lower recoil) kick realizations.
Black solid line: Sgr A* when kicks are excluded. The red thick and thin lines show the 
K1 and K2 realizations favoring the largest black holes. The blue thick and thin lines 
show the K1 and K2 realizations favoring the smallest black holes. Here, the scatter is 
much larger than in the FPL model, particularly for the higher K1 recoil distribution.}

\end{center}
\end{figure}

Gravitational wave recoil matters much more in the PGA model, as can be seen in figure 10 
which shows the central SMBH mass of the Milky Way analogue as a function of redshift. 
Although the final black hole mass lies in a much wider range of masses, these outliers 
in mass are rare. Figure 11 shows the black hole mass with the largest probability at 
redshifts z=3.0 (blue); 1.5(red); and 0.0(black) for PGA models. The black line shows that for the lower 
recoil velocities, K2 case at z=0, the typical black hole mass is in the Sgr A* range 90$\%$ 
of the time. Note that since it is thought that the black hole spin vectors may align with 
one another before coalescence (Bogdanovic et al. 2007, Sperhake 2009, Kesden et al. 2010), 
we consider this our best and most realistic model. If SMBH grows according to the PGA model then
kicks must be in the lower range described by K2 distribution. However, this does imply a scatter in 
the $M$ --$\sigma$ relation at the low mass end.

\begin{figure}
\vspace{5mm}
\begin{center}
\includegraphics [height=4.in,width=2.6in,angle=0]{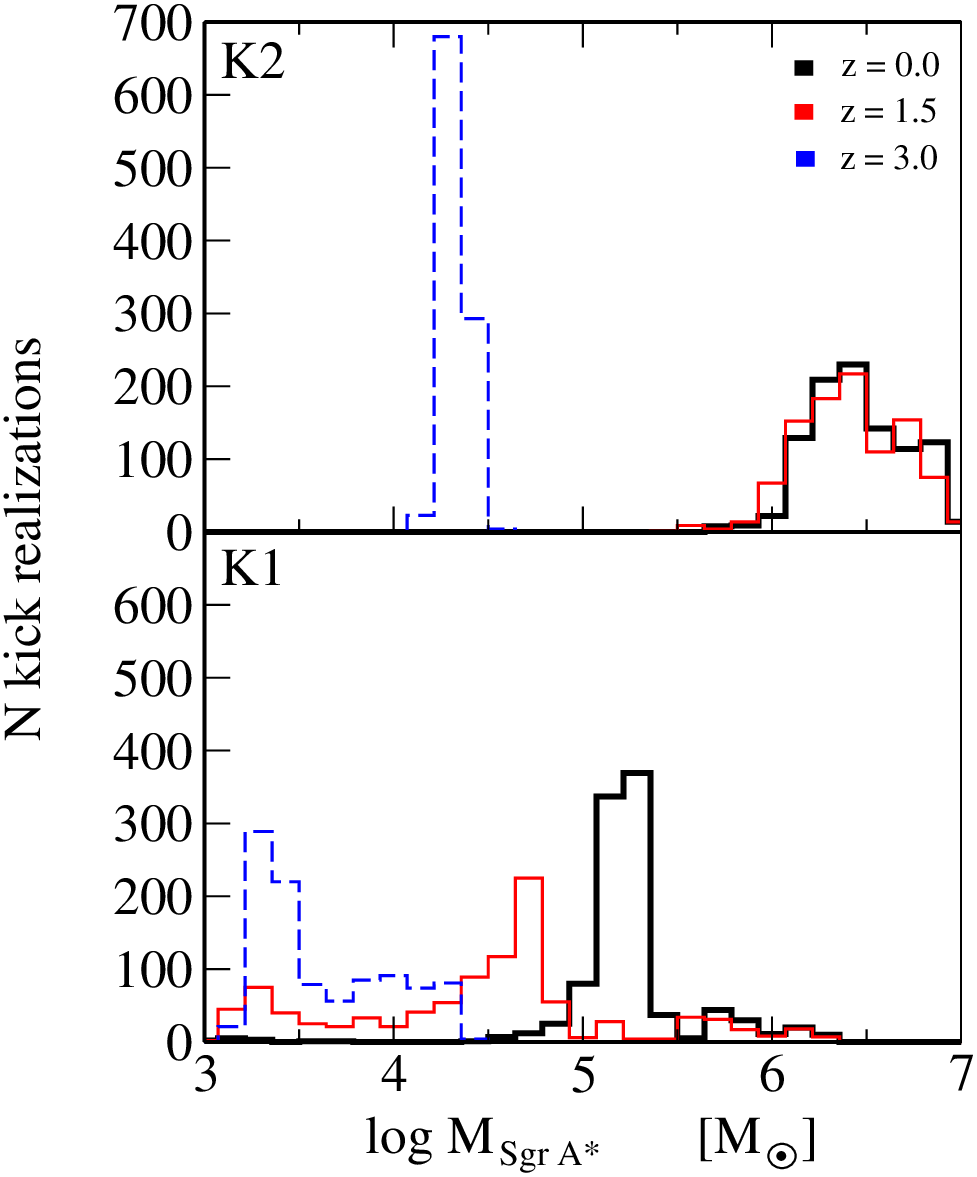}
\caption[Fig 11.]{Upper panel: Number of lower recoil K2 realizations in the PGA model where Sgr A* 
reaches a certain mass at a certain redshift. Bottom panel: Number of K1 realizations in 
the PGA model where Sgr A* reaches a certain mass at a certain redshift. Histograms are 
at z=3.0 (blue dashed); z=1.5 (red thin); and z=0 (thick black). In the lower recoil K2 case, Sgr A* 
reaches the observed value at z=0 in more than 90$\%$ of kick realizations.}
\end{center}
\end{figure}

\section{Other populations of massive black holes}

We distinguish three different massive black hole populations in our merger trees:
black holes at the centers of local field dwarf galaxies (other halos in VL-2 outside of the 
Milky Way halo); rogue black holes scattered through the Milky Way halo (remnants of satellites 
which merge with Milky Way but do not reach the center in less than Hubble time, Micic et al. 2007, 
Bellovary et al. 2010); and black holes ejected from their host halos to occupy the intergalactic medium 
(if the satellite does reach the center of Milky Way, the following black hole merger might lead 
to the black hole ejection).

\subsection{Massive Black Holes in the Local Dwarfs}

For dark matter halo masses above roughly $5 \times 10^{11} \Msun$, the dark matter halo mass 
correlates well with the mass of the supermassive black hole (Ferrarese 2002):

\begin{equation}
{M_{\rm BH}\over {10^8 \Msun}} \sim A \Bigg( {{M_{\rm DMH}} \over {10^{12} \Msun}} \Bigg)^{m},
\end{equation}
where M$_{\rm DMH}$ is the dark matter halo mass, A=0.1 , and m=1.65 defines the slope in the relation.
Below this mass, there is tentative evidence 
that halos are less effective at forming massive black holes, and may even be unable to form 
them (Ferrarese 2002). To be fair, though, if this relation does hold down to a $10^{11} \Msun$ 
halo, the expected $2 \times 10^4 \Msun$ black hole would be difficult to detect, observationally.

\begin{figure*}
\vspace{10mm}
 \epsfig{figure=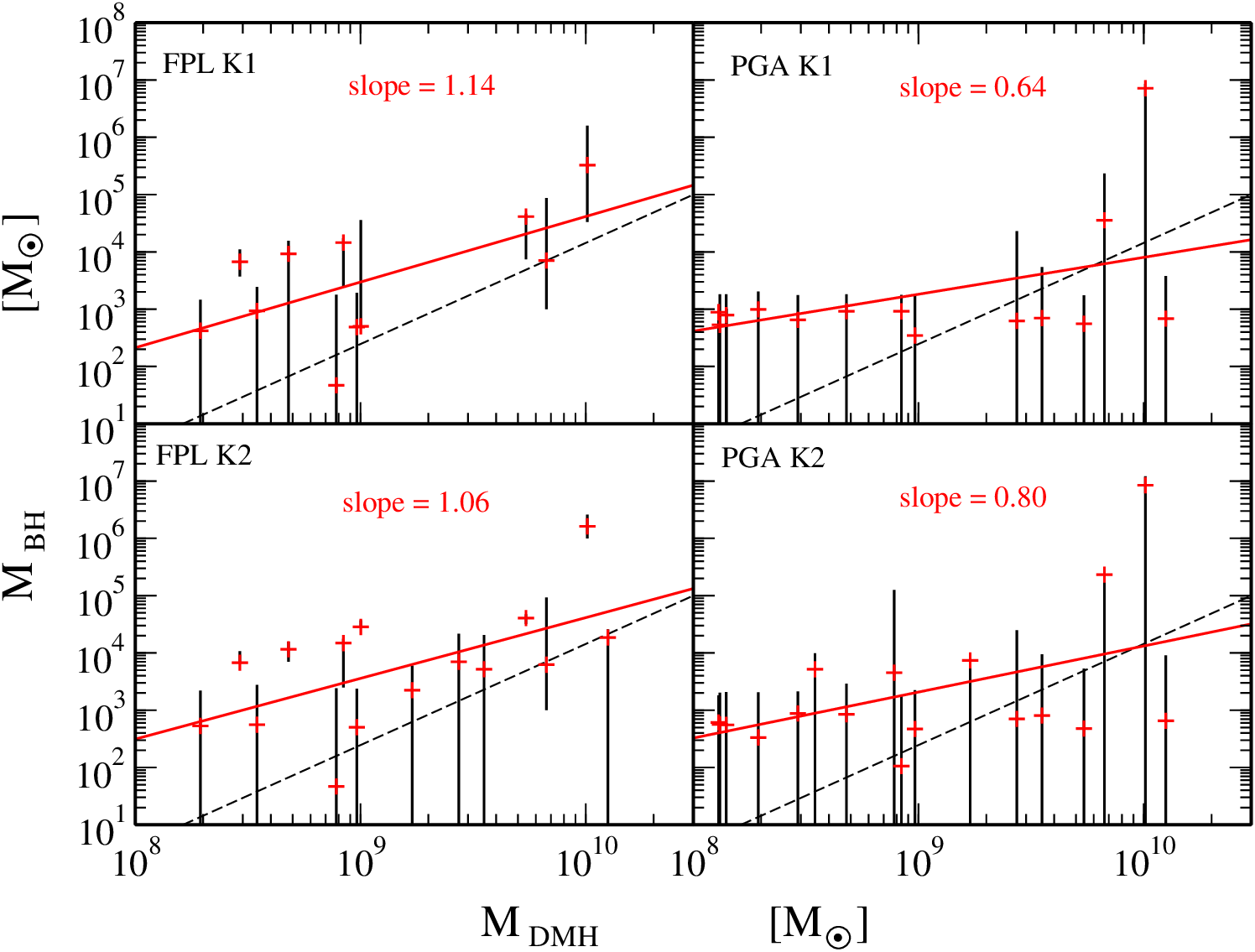, height=0.840\textwidth, width=0.990\textwidth}
 \vspace{5mm}
 \caption{Black hole demography for the FPL K1 model - upper left; FPL lower recoil K2 model 
- bottom left; PGA K1 - upper right; and PGA lower recoil K2 - bottom right. The dashed line in every panel 
represents the Ferrarese relation with a slope = 1.65. Red pluses: the most common central black hole mass. 
For the rest of 35 field dwarf galaxies most common outcome is absence of central massive black hole. 
Red line: linear fit to the most common black hole masses. Black vertical bars: the scatter 
in final black hole mass produced by kicks. This would reflect the scatter in relations that 
connect the central black hole mass with the properties of field dwarf galaxies.}

\end{figure*}

Outside of the Milky Way, there are 35 local field dwarf galaxies in VL-2 identified at z=0 in the mass range 
10$^7\Msun$ - 10$^{11}\Msun$. Figure 12 shows the black hole mass at the centers of these field dwarfs 
as a function of dark matter halo mass in the FPL (left) and PGA (right) models. Thick black vertical 
lines represent minimum and maximum values for the black hole mass in 1,000 kick realizations. The 
Red pluses corresponding to the vertical black lines represent most common mass between minimum and maximum. 
If a red plus is absent, the most common black hole mass for that halo is zero. Linear fits to the most 
common masses are presented by red lines. Note that for our linear fits, we exclude those field dwarfs that have 
ejected their central black holes entirely in more than 50$\%$ of kick realizations. 

The FPL model predicts an $M_{\rm BH}$--$M_{\rm DMH}$ for field dwarf galaxies in K1 and lower recoil 
velocities K2 cases as follows. 

For FPL K1:
\begin{equation}
{M_{\rm BH}\over {10^8 \Msun}} \sim 0.08 \Bigg( {{M_{\rm DMH}} \over {10^{12} \Msun}} \Bigg)^{1.14 \pm 0.50},
\end{equation}
while FPL lower recoil K2 has a slightly shallower slope:
\begin{equation}
{M_{\rm BH}\over {10^8 \Msun}} \sim 0.05 \Bigg( {{M_{\rm DMH}} \over {10^{12} \Msun}} \Bigg)^{1.06 \pm 0.41},
\end{equation}
The PGA K1 model, on the other hand, predicts: 
\begin{equation}
{M_{\rm BH}\over {10^8 \Msun}} \sim 0.002 \Bigg( {{M_{\rm DMH}} \over {10^{12} \Msun}} \Bigg)^{0.64 \pm 0.36},
\end{equation}
and finally, the PGA lower recoil K2 model has the following relation:
\begin{equation}
{M_{\rm BH}\over {10^8 \Msun}} \sim 0.005 \Bigg( {{M_{\rm DMH}} \over {10^{12} \Msun}} \Bigg)^{0.80 \pm 0.36}.
\end{equation}

A generic prediction of each model is that the slope is shallower than the Ferrarese relation 
presented by dashed lines in Fig 12. For FPL K1 and lower recoil K2 models, the slope m is 1.14 and 1.06, close to 
$M_{\rm DMH}$ / $M_{\rm BH}$ = 10$^5$. PGA models predict an even shallower slope of
0.64 in K1 and 0.80 in the case of lower recoil velocities K2. This tests the PGA model by predicting a 
substantially flatter mass function for lower mass SMBH. We do point that these differences between 
FPL and PGA models might be insignificant, considering that scatter around the linear fits is substantial.

\subsection{Rogue Massive Black Holes in the Milky Way}

\begin{figure*}
 \epsfig{figure=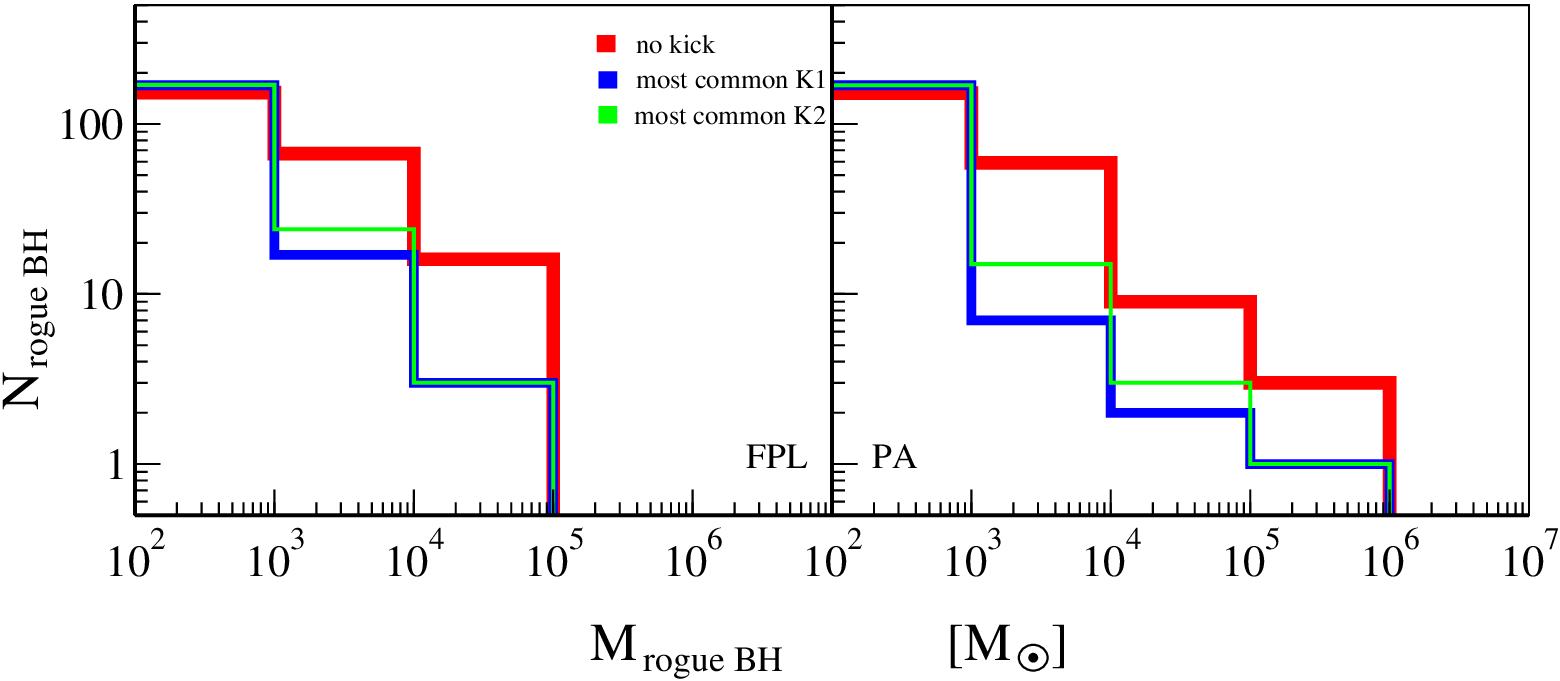, height=0.350\textwidth, width=0.990\textwidth}
 \vspace{5mm}
 \caption{Number of rogue black holes spread through the Milky Way halo as 
a function of mass, for the FPL (left panel) and PGA (right panel) models. 
Red histograms -- the merger tree that excludes black hole kicks. Blue histogram -- 
the most common merger tree in the K1 kick distribution. Green histogram -- the most common merger 
tree for the lower recoil K2 case. These black holes originate from the centers of satellites that 
merged with the Milky Way analogue but had merger timescales larger than a Hubble time. 
As the black holes fail to reach the Milky Way center, they orbit in the Milky Way halo. 
These models predict hundreds of rogue black holes with masses reaching up to $\sim$ 200,000 $\Msun$.}

\end{figure*}

Rogue black holes are carried into the Milky Way halo by their host satellites, but have halo merger time 
scales longer than a Hubble time. Over the span of the simulation, the Milky Way accretes 669 satellite halos
which host a black hole, 
and 249 have orbital decay times that are longer than a Hubble time (see figure 1 and figure 4 for the merger 
time scale, merger rate and mass ratios of these halos). By redshift zero, then, these halos are still orbiting 
at significant distances from the center of Milky Way --  and although the halos themselves have been stripped 
by the primary potential, they still host massive black holes. Figure 13 shows the number of rogue black holes 
as a function of mass for the FPL (left) and PGA (right) models.  As before, the upper limit for black hole mass 
excludes kicks and is presented in red in both panels. For each kick model, we also present the number of rogue 
black holes with the typical mass over the realizations. The PGA model predicts that there should be at least
one massive rogue black hole in the Milky way halo with mass in range 10$^5\Msun$ -10$^6\Msun$.

\begin{figure}
\vspace{5mm}
\begin{center}
\includegraphics [width=2.5in,angle=0]{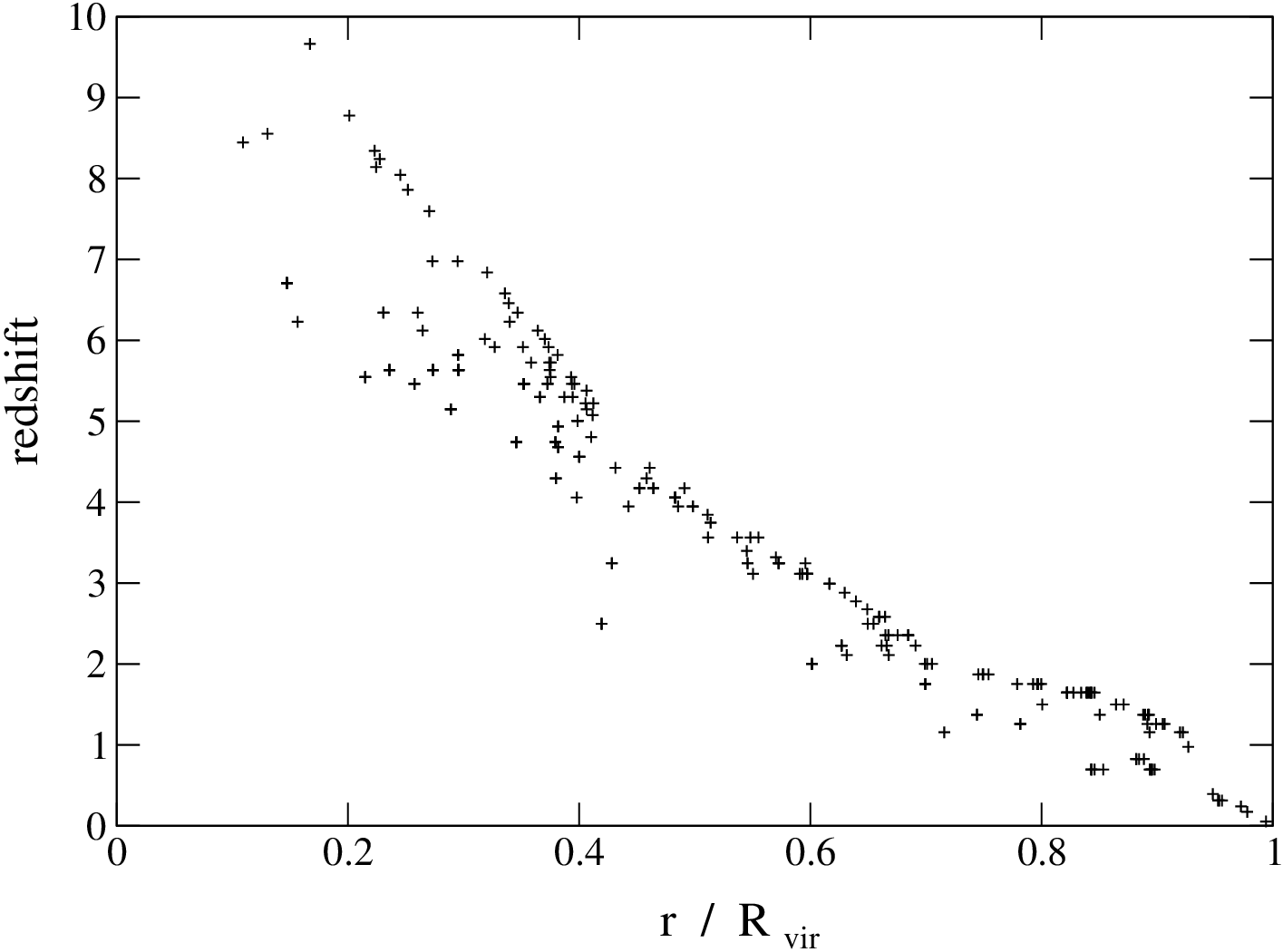}
\caption[Fig 14.]{Redshift of satellite mergers with the Milky Way analogue as a function
of the current position in the Milky Way halo. Those satellites that merged with
the Milky Way at high redshift have sunk closer to the Galactic center. In later mergers, the
satellites have just entered the Milky Way and are closer to the virial radius. Since
each satellite potentially carries a black hole, this figure shows the distance
of rogue black holes from the center of the Milky Way.}
\end{center}
\end{figure}

The distance from the Milky Way center in all models is correlated with the redshift at which galaxies merge 
(figure 14). If the merger occurred early at high redshift, the black hole had more time to sink to the center, 
and for mergers close to z=0, the black hole has just entered Milky Way halo. The redshift at which the halo 
merges with the Milky Way also marks the end of the rogue black hole growth. Hence, rogue black holes that enter 
the Milky Way at later times have had more time to grow, and will be more massive than those accreted through 
early mergers. Figures 15 and 16 show how rogue black hole mass, satellite merger redshift, and distance from 
the Milky Way are correlated. Figure 15 represents the FPL model and figure 16 represents the PGA model. 
K1 realizations are shown in the left panels, and lower recoil K2 realizations in the right.

\begin{figure*}
\vspace{10mm}
 \epsfig{figure=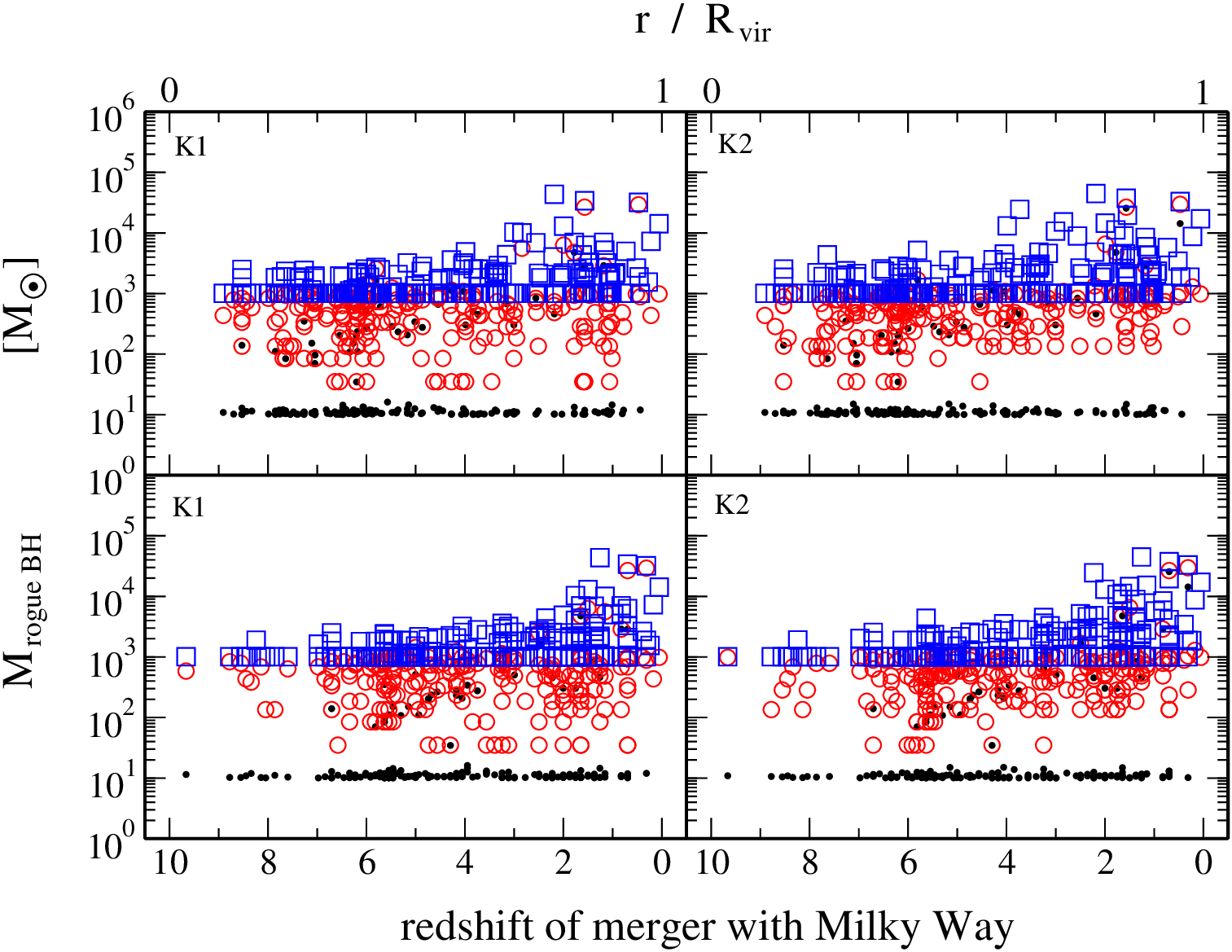, height=0.9900\textwidth, width=0.990\textwidth}
 \vspace{5mm}
 \caption{Rogue black hole mass as a function of Galactic distance and progenitor halo 
redshift for the FPL model. Left panels: the K1 distribution. Right panels: the lower recoil K2 case.
The blue squares represent the most massive rogue black holes; red circles represent the typical 
black hole mass, and black dots represent the lightest. Those black holes carried by satellites 
that merge later have had more time to grow, so the rogue black hole mass correlates with both 
distance from the center and merger redshift. The most massive rogues reach $\sim$ 30,000 $\Msun$ 
at 0.8 -- 1.0 of virial radius. }

\end{figure*}

\begin{figure*}
\vspace{10mm}
 \epsfig{figure=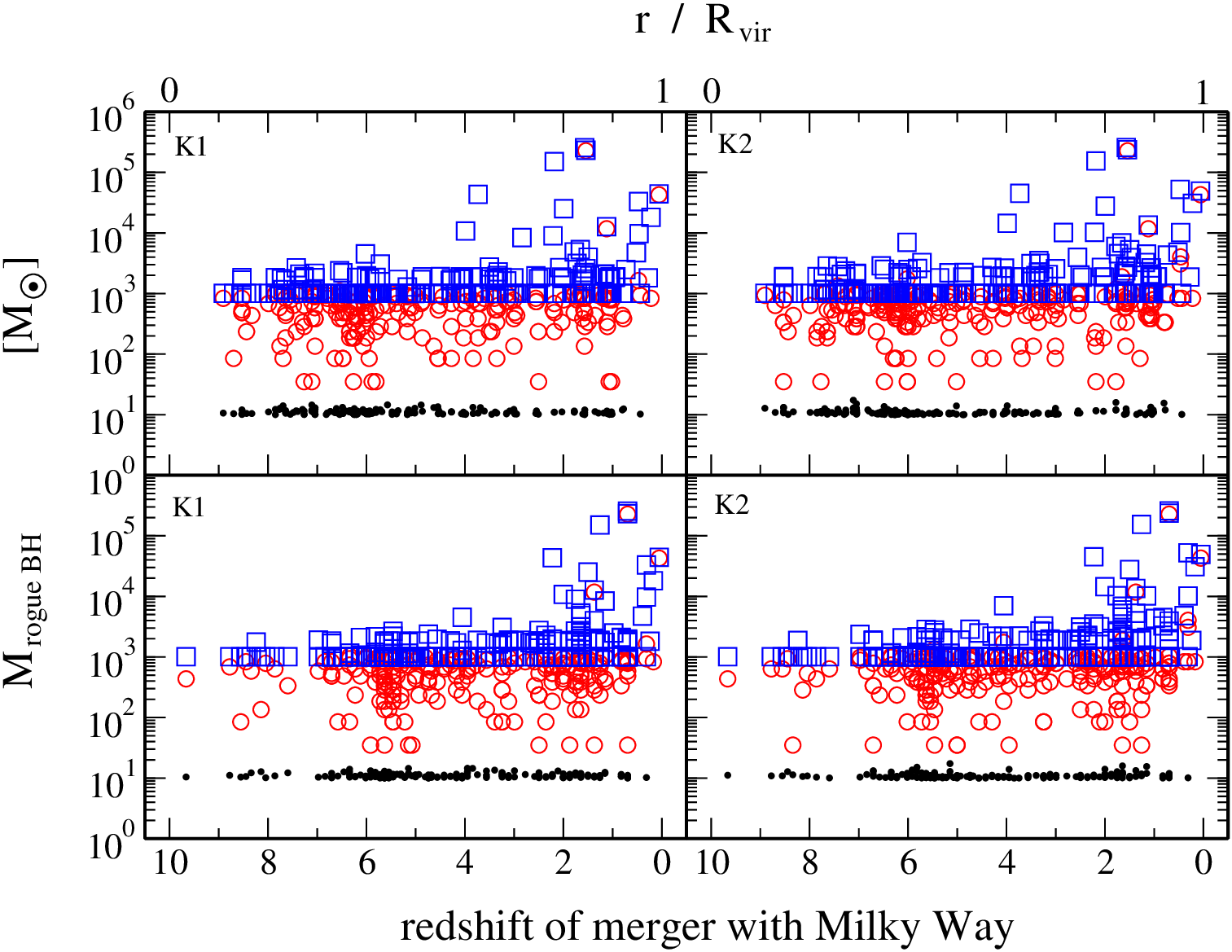, height=0.9900\textwidth, width=0.990\textwidth}
 \vspace{5mm}
 \caption{Rogue black hole mass as a function of Galactic distance and progenitor halo 
redshift for the PGA model. Left panels: the K1 distribution. Right panels: the lower recoil K2 case.
The blue squares represent the most massive rogue black holes; red circles represent the typical black 
hole mass, and black dots represent the lightest. The most common rogue black hole mass remains at the 
initial black hole seed mass, but a number of massive rogue black holes are greater than 10,000 $\Msun$.
The most massive is M$_{\rm rogue} \sim 230,000\Msun$ at 225 kpc from the Galactic Center (0.8 -- 0.9 of
virial radius).}

\end{figure*}

Unfortunately, both figures clearly show that the most massive rogue black holes
are furthest away and may be very difficult to observe. We estimate the Bondi accretion onto these
black holes to determine how observable they may be and we describe the process below.

There is sufficient evidence that the halos of galaxies, from massive ellipticals to isolated 
spirals, can host diffuse hot halo gas (Matthews $\&$ Brighenti 2003, O'Sullivan et al. 2001, 
Pedersen et al. 2006). As these black holes orbit within the primary halo, they can accrete 
from this ambient hot halo gas via Bondi-Hoyle accretion. Here, the mass accretion rate can 
then be described by:

\begin{equation}
{{\dot M_{\rm BH}}} = {{4 \pi G^2 M_{\rm BH}^2 \rho_{\rm ISM}} \over { (c_s^2 + v^2)^{3/2}}},
\label{eqn:Bondi}
\end{equation}

\noindent where $\rho_{\rm ISM}$ is the density of the halo gas, $c_s$ is the gas sound speed, 
and $v$ is the velocity of the black hole. For these rogue black holes, we add this more 
quiescent form of accretion from the time the black hole has entered the virial radius of the 
halo in order to determine whether any may be visible today. To model 
the hot halo gas, we use a technique outlines in Sinha \& Holley-Bockelmann (2009). In short, 
we assume it has an isothermal density profile consistent with X-ray 
observations of halo gas in ellipticals (Matthews $\&$ Brighenti 2003), and that the ideal gas 
is in hydrostatic equilibrium with the gravitational potential of the halo. We assume a gas core 
radius of $0.3 {\rm R}_s$, where R$_s$ is the scale radius of the NFW halo, and that the gas 
temperature is the virial temperature at the virial radius. This yields a central gas 
temperature of $\sim 2 \times 10^6 K$ for a dark matter halo of mass $2 \times 10^{12} {\rm M}_\odot$, 
which is also consistent with observations (Pedersen et al. 2006). At redshift zero, we assume 
the black holes are on bound elliptical orbits of eccentricity $0.8$ to match simulation 
predictions for satellite mergers (e.g. Ghigna et al. 1998, Sales et al. 2007).

\begin{figure*}
\vspace{10mm}
 \epsfig{figure=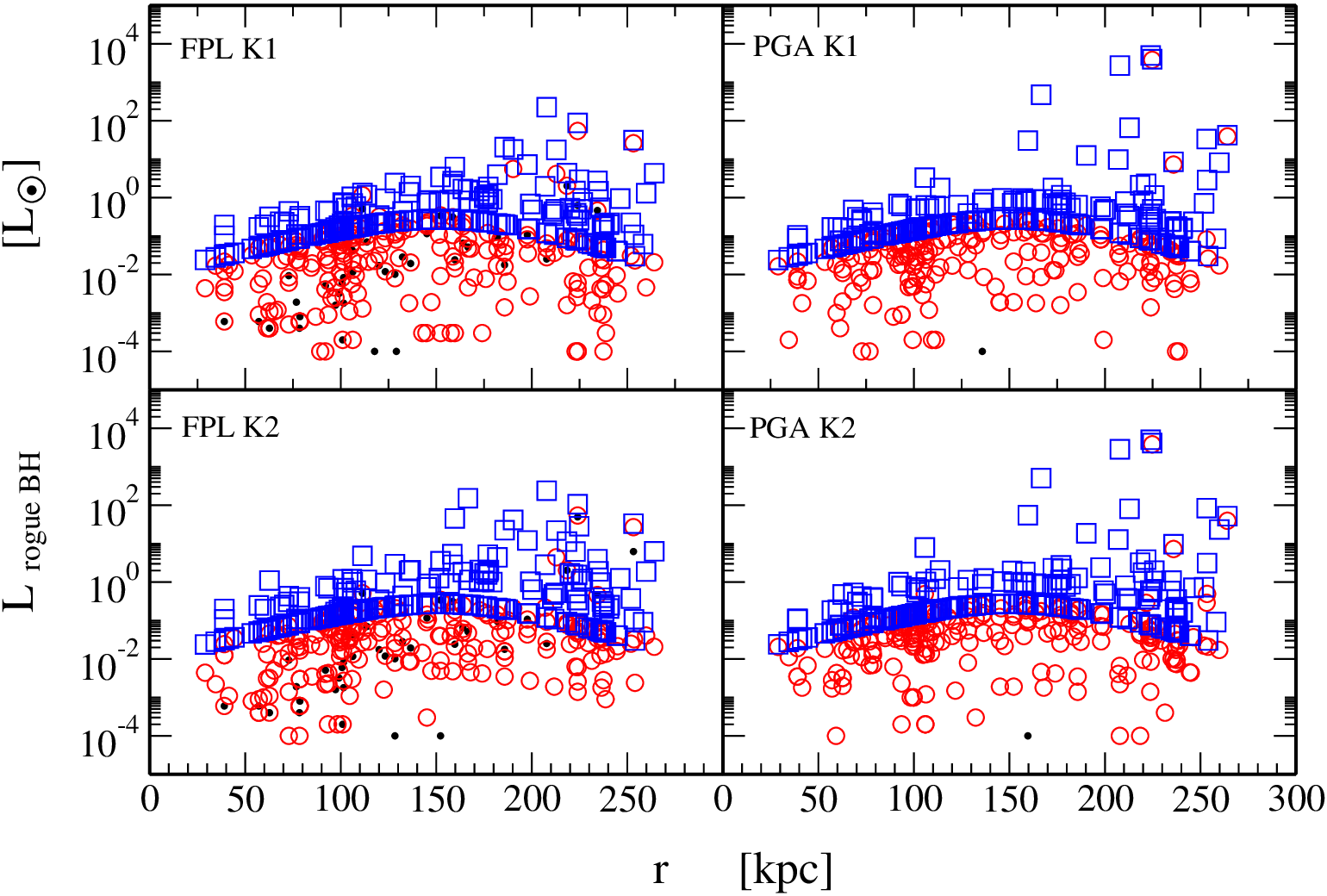, height=0.9900\textwidth, width=0.990\textwidth}
 \vspace{5mm}
 \caption{Bolometric luminosity of rogue black holes as a function of distance 
from the primary halo center for the FPL (left) and PGA (right) models. The K1 case is on 
the upper panels and the lower recoil K2 case on the bottom. Blue squares represent the most massive 
rogue black holes; red circles represent the typical black hole mass, and black dots represent 
the smallest. Black holes are assumed to accrete via a Bondi-Hoyle mechanism 
from the ambient gas in the Milky Way. Most rogue black holes are well below solar 
luminosity, but a few are expected to be luminous Xray sources (just under the ULX cut-off) 
that reside in the outskirts of the Milky Way halo. The most massive of them is in PGA model and has 
a luminosity of 4000 $ {\rm L}_\odot$ .}

\end{figure*}

\begin{figure*}
\vspace{10mm}
 \epsfig{figure=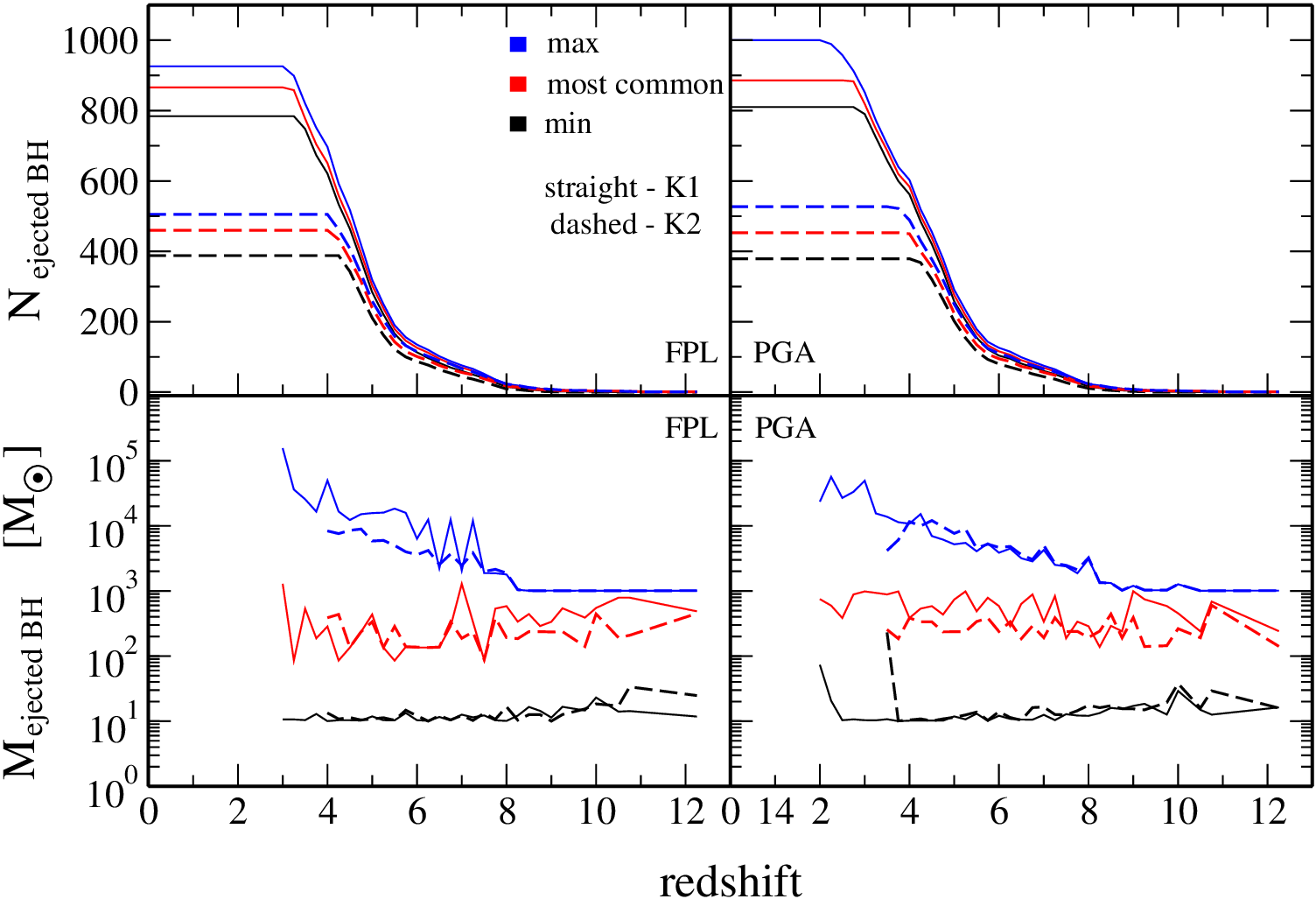, height=0.9900\textwidth, width=0.990\textwidth}
 \vspace{5mm}
 \caption{Upper panels: The cumulative number of ejected black holes as a 
function of redshift for all halos in VL-2 for the FPL (left) and PGA (right) models. 
The K1 (solid line) and lower recoil K2 (dashed line) kick distributions are shown. In blue is the 
maximum cumulative number of ejected black holes; in red is the typical number; 
in black -- the minimum. Lower panels: The black hole mass of the ejected population. 
The colors are the same as in the top panels.}

\end{figure*}

By redshift zero, these black holes will have grown to a new mass, M$_{\rm rogue}$, and will have 
a new accretion rate from equation~\ref{eqn:Bondi}. The radiated luminosity from this process 
can be determined from L$_{\rm rogue} = \eta {\dot {\rm M}_{\rm BH}} c^2$, where $\eta$ 
describes how efficiently mass is converted to energy; the standard assumption for an 
accretion disk around a spinning black hole sets $\eta\sim 0.1$ (Shakura $\&$ Syunyaev 1973). 
We find that our most massive rogue black hole M$_{\rm rogue}$ = 227,643 $\Msun$, residing at 225 kpc 
from the Galactic center, can radiate at 3,834 $ {\rm L}_\odot$ (figure 17, PGA panels on the right), 
making it somewhat fainter than the brightest ULXs; the fact that the hot halo 
gas is so tenuous at these radii is what accounts for the relatively low luminosity. 

However, if these rogue black holes are common, they will be observable in the outskirts of the Milky 
Way halo with Chandra, and the most luminous can have sufficient signal-to-nose for spectral 
characterization. Note, though, that a significant fraction of these may have advection-dominated 
accretion flows, rather than thin accretion disks, and hence $\eta$ may be much less than 0.1; 
this would naturally make the black hole less luminous, but with a harder spectral signature. 
It has also been suggested that rogue black holes in the molecular or cold neutral part of the 
ISM should be more easily detected in the radio than in X-rays (Maccarone 2004, Maccarone et al. 2005). 
Maccarone 2005, for example, calculates that a 2,600 $\Msun$ black hole should be observable with 
LOFAR (above 30$\mu$Jy) in the radio out to about 15 kpc -- much closer than the most massive 
rogue black hole in our model. Assuming larger black hole velocities 
and a more efficient Bondi accretion rate, rogue black holes may be observed at even larger distances. 
Even if they are not observable electromagnetically, they may be detected by mesolensing (di Stefano 2007).

\subsection{Ejected Massive Black Holes}

As the SMBH grows, incoming black holes are kicked from the center of the host halo, forming a population 
of ``ejected'' black holes. Figure 18 shows the cumulative number of black holes lost from the merger tree 
due to kicks larger than the escape velocity in the top panels, and the ejected black hole mass as a function 
of redshift on the bottom panels. Again, we show the FPL model on the left and our PGA model on the right. 
The redshift of the last SMBH ejection is between z=2-3 for the PGA model, and between z=3-4 for the FPL model. 
When we compare the difference in total numbers of ejected BHs between K1 and lower recoil K2 realizations 
(upper panels in figure 18), both the FPL and PGA model predict the the K1 case ejects up to 50$\%$ more black 
holes. The average masses of ejected black holes are similar in the FPL and PGA models (bottom panels in 
figure 18) with the most common ejected black hole of 100 - 1,000 $\Msun$.

The mass ratio of merging black holes sets the kick amplitude, with large differences between 
the incoming black hole mass creating a small kick. The typical number of ejected black holes 
per realization as a function of mass ratio is presented in figure 19. The FPL and PGA distributions 
are again very similar regardless of the kick distribution. The main difference is that higher mass 
ratio encounters are more common in the PGA K1 model. Since the growth of black holes in the FPL model 
is very efficient -- at Eddington rate -- the incoming black holes grow faster and this decreases the 
mass ratio when the merger occurs. Hence, there are more mergers at q $\leq$ 0.3 in FPL K1 and 
more mergers at q $\geq$ 0.3 in PGA K1. 

Some of the ejected massive black holes might contribute to the rogue black hole 
population. Micic et al. 2006 has shown that high redshift ejections of black holes
from their host halos in the Local Group type environment could lead to their 
capture by Milky Way potential. Tracking these black holes are beyond the scope of this 
paper and will be addressed in future work.

\begin{figure}
\vspace{5mm}
\begin{center}
\includegraphics [width=3.in,angle=0]{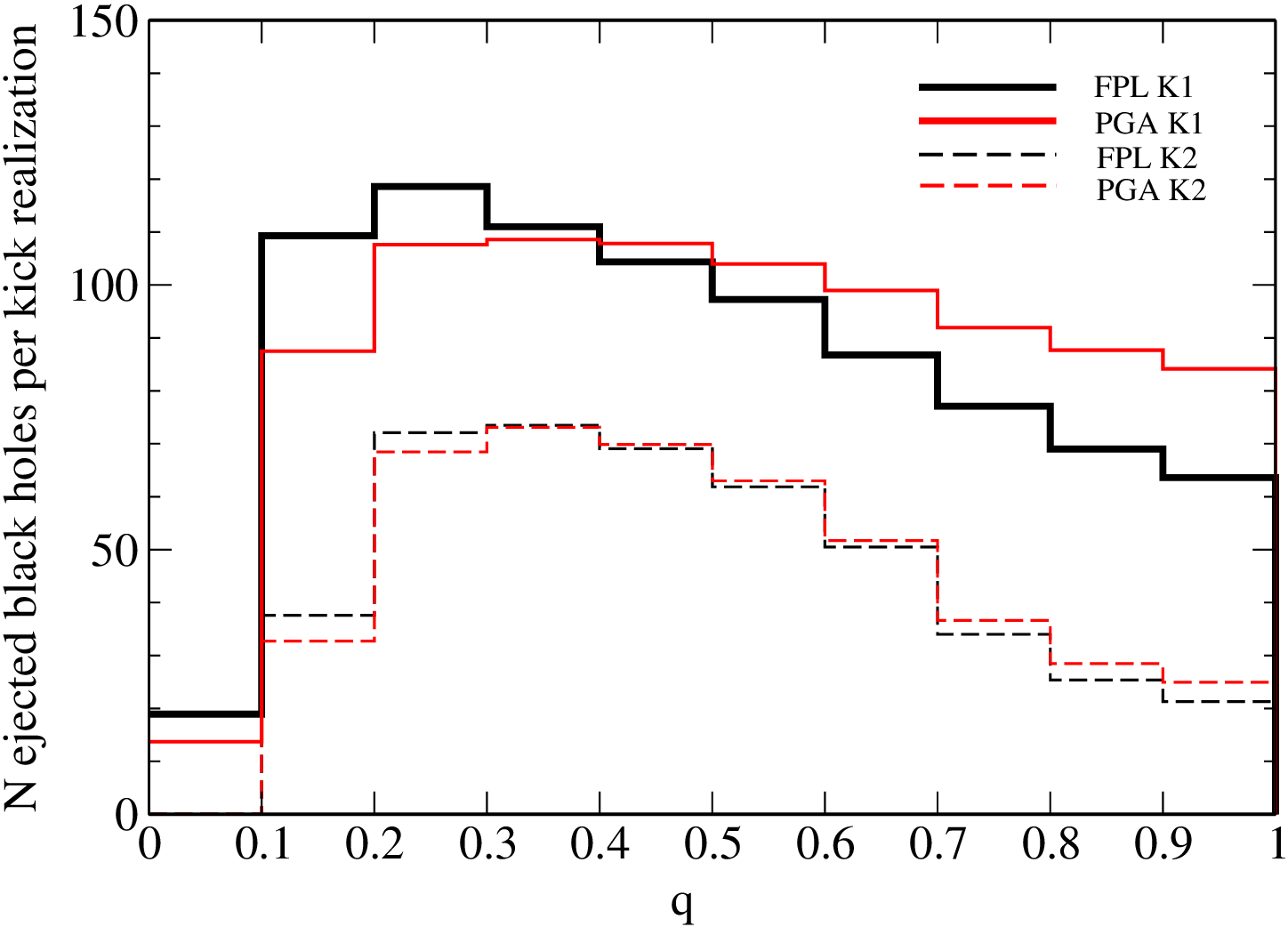}
\caption[Fig 19.]{Number of ejected black holes per realization as a function
of the mass ratio of the merging black holes. Black lines represent the FPL model, 
and red lines represent the PGA model. Thick likes show the K1 distribution, while 
thin dashed lines present the lower recoil K2. Since the K1 model yields on average an order of 
magnitude larger kick velocities than K2 lower recoil model, the number of ejected black holes is larger for K1. }
\end{center}
\end{figure}

\section{DISCUSSION AND CONCLUSIONS}

Using a very high resolution small volume cosmological N-body simulation (Via Lactea II), 
we construct the merger history of black holes in low mass dark matter halos in a Local 
Group analog. We found that the method used to estimate the dynamical friction timescale 
makes a large difference in the total rate and redshift distribution of black hole mergers 
(see also Holley-Bockelmann et al. 2010). In particular, for example, an N-body based 
dynamical friction estimate (Boylan-Kolchin 2008) yields longer merger timescales than 
does Chandrasekhar dynamical friction. These larger, more realistic merger timescales will 
postpone black hole mergers to lower redshifts. Moreover, if most of the black hole growth 
is tied to gas accretion that is activated by major mergers, then using Chandrasekhar 
dynamical friction will overestimate the black hole mass.

We studied the growth and merger rate of black holes from seeds at redshift $\sim$ 20 to 
low mass SMBHs at redshift zero, analytically incorporating the important sub-resolution 
physics to model the black hole dynamics and gas accretion. We find that having a constant 
mass black hole seeds, or a narrow BHMF leads to nearly equal mass ratio black hole mergers. 
These mergers have largest gravitational wave recoil velocities, which eject the black holes 
from their host galaxies and suppresses SMBH growth. A wider range BHMF presents far fewer 
problems in growing SMBHs, even when the black holes spins are large. 

Many groups are beginning to explore the formation and evolution of massive supermassive 
black holes in large volume cosmological volumes, with a sophisticated and self-consistent 
treatment of the sub-grid physics (e.g. Croton et al. 2006, Sijacki et al. 2007, di Matteo et al. 2008, 
Sijacki et al. 2009). With larger 
simulation volumes, these simulations are capable of forming hundreds of massive SMBHs at 
the centers of massive elliptical galaxies and clusters of galaxies. In fact, the FPL model 
we adopt is a semi-analytic model (Somerville et al. 2008) resulting from these techniques. 
Due to high computational cost, however, the same numerical approach has not been applied 
in smaller cosmological volumes where the bulk of the halo evolution is non-linear. Here, 
high redshift dwarf galaxies merge to form the Local Group galaxies, and massive black holes 
have masses well below 10$^7\Msun$.  

We use small volume nbody simulations and apply two semi-analytic recipes for SMBH growth. 
One class of models is always limited by the black hole fundamental plane (FPL). Here, the 
growing black hole mass is a function of the total stellar mass in the galaxy bulge. This 
approach has successfully produced SMBHs in massive galaxies. We show that the FPL model
can also be used for black hole growth in small volumes. In particular, black hole mergers 
combined with a) gas accretion in a quasar mode and b) a time averaged Bondi-Hoyle accretion 
rate of 7 $\times10^{-4}\Msun$yr$^{-1}$ in a radio mode, can produce the observed mass of the 
Sgr A* SMBH. Today, the observed Sgr A* accretion rate is an order of magnitude less efficient 
than this radio mode accretion rate, but this may not be unreasonable, since the Milky Way is 
not considered an AGN anymore. Other low mass AGNs in the local Universe are observed to have 
10 - 100 times more efficient gas accretion. The FPL model, by design, guarantees the formation 
of a fundamental plane black hole. However, in reality, not every galaxy has such a system, 
due to the competition between black hole growth and nuclear star cluster formation in low mass 
spheroids. 

We have proposed an alternative method which could provide deeper insight into the formation of 
the lightest supermassive black holes ($\leq$10$^7\Msun$) by incorporating growth prescriptions 
calculated in small scale merger simulations. Gas accretion onto the black hole is always suppressed 
by stellar or AGN feedback, and depends on the structure of the gas, as well as the microphysics 
(such as turbulence, fully relativistic magnetohydrodynamics, etc) that is currently physically 
beyond modeling. Our model provides a framework into which the microphysics can easily be 
incorporated once simulations and observations reveal the nature of these processes. In contrast, 
in the FPL model, these effects are irrelevant, since gas accretion occurs in the same manner in all 
$\leq$ 10:1 galaxy mergers. The final outcome is always a fundamental plane black hole since feedback 
rapidly drives the evolution of fundamental plane rather than the growth of black holes. In the PGA 
model, we can incorporate an accretion efficiency parameter into the gas accretion prescription to 
make black hole growth less efficient for stronger feedback and vice versa. We use a formalism that 
relates the gas accretion efficiency to galaxy merger ratios; gas accretion is most efficient for 
major mergers and negligible for minor mergers (Cox et al. 2008).  In addition, minor mergers will 
last much longer than equal mass mergers. When these two effects combine, the PGA model produces 
short bursts of high rate gas accretion for equal mass mergers, and long, slow gas accretion for 
10:1 mass ratio mergers. As a  result, most black holes in the PGA model never reach the Eddington rate.

We also modeled the effect of gravitational wave recoil on assembling the lightest SMBHs. One way 
to mitigate gravitational wave recoil is to increase the black hole merger symmetry; smooth dense 
gas, for example, may quickly align the spin vectors of each SMBH before coalescence (Bogdanovic 
et al. 2007, Dotti et al. 2010). In order to understand how damaging gravitational wave recoil can 
be, and to pin down the epoch that a growing black hole is most vulnerable to recoil, we do not 
pre-select the merger configurations with the lowest recoils in K1 and select spins aligned to the 
orbital angular momentum in lower recoil model K2. In previous work, when black holes are merging in 
massive stellar systems, the large gravitational potential creates large escape velocities. Hence, a K2 kick 
distribution with much lower recoil velocities helps retain the central black holes (Volonteri et 
al. 2010). In our merger tree, most of the dark matter halos have masses below 10$^{10}\Msun$, and 
the typical halo mass is 10$^7\Msun$ - 10$^9\Msun$. Since the escape velocities in these systems 
are below 100 $\kms$, even the moderate kicks will eject the SMBH. When gravitational recoil is 
included in the FPL model, the final black hole mass is unaffected in most cases. This is the 
consequence of the FPL requirement of a very high gas accretion efficiency. Even in the most damaging 
K1 realization where the black hole is ejected twice from the merger tree, the accretion of gas is so 
efficient that a $\sim 100\Msun$ black hole that is ushered to the center has time to grow to 
2$\times$10$^6\Msun$. When gravitational recoil is included in the PGA model, the final mass of the 
central black hole in the K1 realizations is lower by roughly an order of magnitude than in K2 (lower 
recoil model). 90 $\%$ of the K2 realizations are in the Sgr A* mass range. PGA model predicts that kicks have
to follow lower distribution of gravitational recoil velocities, K2.

When we look at the mass of the black holes at the centers of local field dwarf galaxies, none of the 
models, FPL or PGA, matches the extrapolation of the Ferrarese (2002), relation which relates black 
hole masses to the masses of host dark matter halos. Even when we remove gas accretion as a mechanism 
for black hole growth, their masses will be larger than predicted by Ferrarese relation as the result
of seed black hole mergers, despite the fact that the initial black hole seeds were Pop III remnants 
and therefore considered light seeds. It is possible that the black hole fundamental 
plane exists for low mass SMBHs in field dwarf galaxies, but it may simply have a different slope than for high 
mass SMBHs; indeed this was alluded to in the original Ferrarese work. We do find a shallower slope in 
all models. In FPL, the slope is close linear and tracks the following proportion:
M$_{\rm DMH}$ / M$_{\rm BH}$ = 10$^5$.  The PGA model predicts an even shallower slope --  
0.64 - 0.80 -- although the scatter caused by gravitational wave recoil
makes it hard to distinguish between FPL and PA. Note that gas consumption 
in field dwarf galaxies may favor nuclear star cluster formation rather than a massive black hole. 
Both FPL and PGA models are consistent with this picture since roughly half the dwarfs contain no central 
black hole, though the 
PGA models predict more empty halos on average than FPL.
It is important to note that the black hole demographics in the dwarf galaxy regime is critical to 
distinguish between the FPL and PGA models.

We also identified two massive rogue black hole populations. First, we find ejected black holes, or 
black holes with kick velocities larger than the host halo escape velocity. We will address the 
properties of this population with a detailed treatment of their dynamics in the future. The second 
are rogue black holes from the remnants of satellites that had merger timescales longer than a Hubble 
time. We find hundreds of these rogue black holes at distances of $\sim$ 200 kpc from the galactic 
center and with masses from 10 - 10$^5\Msun$ (depending on the model). The most massive of these may 
be detected in X-ray and are possible mesolensing candidates. These two populations of black holes, 
which have quite different formation mechanisms, might not be easily disentangled. It is possible 
that a growing primary halo may recapture black hole ejected by recoil, in which case both 
populations occupy a similar parameter space (Micic et al. 2006).

In conclusion, our model successfully reproduces Sgr A* without limiting the black hole growth to 
the black hole fundamental plane. It predicts that kicks have to follow lower recoil velocity distribution (K2). 
It also predicts a substantially flatter mass function (slope = 0.64 -- 0.8) for lower mass 
SMBH in local field dwarfs. It predicts Milky Way rogue black holes, and a population of ejected black holes. 
This model provides cosmological framework into which the results of small scale simulations can easily be 
incorporated in the future. At this point, the model is limited only by the lack of understanding 
of how various feedback mechanisms couple with each other and with the gas accretion onto the black 
hole. There is ample ground to cover in this respect. As one example, Ostriker et al. (2010) claim 
that all previous studies of SMBH growth may overestimate the final black hole mass by $\sim$ two 
orders of magnitude by treating only the energy part of AGN feedback, neglecting the importance of 
momentum and mechanical feedback. Hopefully, this will be resolved in the future.

One issue is how well the VL-2 simulation represents the Milky Way merger history. The reality is 
that VL-2 represents one out many possible evolutionary histories of the Local Group, and not 
necessarily the correct one. The fact that only major mergers activate the quasar mode means that 
the number of major mergers each galaxy goes through has a huge impact on the black hole accretion 
history. This underlines the importance of approaching the problem of Sgr A* growth statistically 
by simulating hundreds of highly resolved small simulation volumes. To resolve thousands of Milky 
Way mass halos with VL-2 resolution, gas physics, and black hole dynamics implemented on the fly 
will be computationally too expensive in the near future.

\section*{ACKNOWLEDGMENTS}

MM acknowledges support from ARC DP DP0665574. KHB acknowledges the support of NSF Career Grant
AST-0847696, as well as the supercomputing support of Vanderbilt's Advanced Center for Computation 
Research and Education, and NASA's Pleiades and Columbia clusters. This project benefited from 
interactions during the Aspen Center for Physics Summer Program.


\begin{thebibliography}{dw}

\bibitem[Aguilar \& White(1986)]{1986ApJ...307...97A} Aguilar, L.A., White, S.D.M., 1986, ApJ, 307, 97 
\bibitem[Alexander et al. 2005]{}Alexander, D.M., Smail, I., Bauer, F.E., 
Chapman, S.C., Blain, A.W., Brandt, W.N., Ivison, R.J., 2005, Natur, 434, 738A
\bibitem[Abel et al. 2000]{Abel2000}Abel, T., Bryan, G., Norman, M., 2000, ApJ, 540, 39     
\bibitem[Abel et al. 2002]{Abel2002}Abel, T., Bryan, G., Norman, M., 2002, Sci, 295, 93A   
\bibitem[Adams et al. 2001]{}Adams, F.C., Graff, D.S., Richstone, D.O., 2001, ApJ, 551L, 31A
\bibitem[Baes et al. 2003]{}Baes, M., Buyle, P., Hau, G.K.T., Dejonghe, H., 2003, MNRAS, 341L, 44B
\bibitem[Baker et al. 2007]{}Baker, J. G., et al. 2007, ApJ, 668, 1140B
\bibitem[Begelman et al. 1980]{1980Natur.287..307B} Begelman, M.C., Blandford, R.D., Rees, M.J., 1980, Natur, 287, 307 
\bibitem[Begelman \& Rees 1978]{}Begelman, M.C., Rees, M.J., 1978, MNRAS, 185, 847B
\bibitem[Bellovary et al. 2010]{}Bellovary, J, M., Governato, F., Quinn, T.R., Wadsley, J., Shen, S., Volonteri, M., 
2010, MNRAS, 721L, 148B
\bibitem[Bender et al. 2005]{}Bender, R., Kormendy, J., Bower, G., 2005, ApJ, 631, 280
\bibitem[Berczik et al. 2006]{}Berczik, P., Merritt, D., Spurzem, R., Bischof, H.P., 2006, ApJ, 642L, 21B   
\bibitem[Binney \& Tremaine 1987]{}Binney, J., Tremaine, S., 1987, Galactic Dynamics (Princeton: Princeton Univ. Press)   
\bibitem[Bogdanovic et al. 2007]{}Bogdanovic, T., Reynolds, C.S., Miller, M.C.,  2007, ApJ, 661L, 147B
\bibitem[Boker et al. 2002]{}Boker, T., Laine, S., van der Marel R.P., Sarzi, M., Rix, H.-W., \&
Ho, L.C., Shields, J.C., et al., 2002, AJ, 123, 1389 
\bibitem[Booth \& Schaye 2009]{}Booth, C. M., Schaye, J., 2009, MNRAS, 398, 53B 
\bibitem[Boylan-Kolchin et al. 2008]{}Boylan-Kolchin, M., Ma, C.P., Quataert, E., 2008, MNRAS, 383, 93B   
\bibitem[Bromm et al. 1999]{1999ApJ...527L...5B}Bromm, V., Coppi, P.S., Larson, R.B., 1999, ApJ, 527, L5 
\bibitem[Bromm \& Loeb 2003]{}Bromm, V., Loeb, A., 2003, ApJ, 596, 34B
\bibitem[Bromm \& Loeb 2004]{}Bromm, V., Loeb, A., 2004, NewA, 9, 353B    
\bibitem[Bullock et al. 2001]{}Bullock, J.S., Kollat, T.S., Sigad, Y., Somerville, R.S., 
Kravtsov, A.V., Klypin, A.A., Primack, J.R., Dekel, A., 2001, MNRAS, 321, 559B   
\bibitem[Burkert \& Silk 2001]{}Burkert, A., Silk, J., 2001, ApJ, 554L, 151B
\bibitem[Campanelli et al. 2007]{}Campanelli, M., Lousto, C.O., Zlochower, Y., 
Merritt, D., 2007, gr.qc, 2133C   
\bibitem[Cattaneo et al. 1999]{}Cattaneo, A., Haehnelt, M.G., Rees, M.J., 1999, MNRAS, 308, 77C
\bibitem[Colpi et al. 1999]{}Colpi, M., Mayer, L., Governato, F., 1999, ApJ, 525, 720C   
\bibitem[Colpi et al. 2007]{} Colpi, M., Dotti, M., 
Mayer, L., \& Kazantzidis, S.\ 2007, ArXiv e-prints, 710, arXiv:0710.5207 
\bibitem[Cote et al. 2006]{}Cote, P., et al., 2006, ApJS, 165, 57
\bibitem[Cox et al. 2008]{}Cox, T. J., Jonsson, P., Somerville, R. S., Primack, J. R., Dekel, A., 2008, MNRAS, 384, 386 
\bibitem[Croton et al. 2006]{}Croton, D.J, Springel, V., White, S.D.M., De Lucia, G., 
Frenk, C.S., Gao, L., Jenkins, A., Kauffmann, G., Navarro, J.F., Yoshida, N., 2006, MNRAS, 365, 11C
\bibitem[David et al. 1987]{}David, L.P., Durisen, R.H., Cohn, H.N., 1987, ApJ, 313, 556D
\bibitem[Dekel et al. 2009]{}Dekel, A., et al. 2009, Natur, 457, 451D
\bibitem[Diemand et al. 2007]{}Diemand, J., Kuhlen, M., Madau, P., 2007, ApJ, 667, 859D 
\bibitem[Diemand et al. 2008]{}Diemand, J., Kuhlen, M., Madau, P., Zemp, M., Moore, B., 
Potter, D., Stadel, J., 2008, Natur, 454, 735D
\bibitem[Di Matteo et al. 2003]{}Di Matteo, T., Croft, R.A.C., Springel, V., 
Hernquist, L., 2003, ApJ, 593, 56D
\bibitem[Di Matteo et al. 2005]{}Di Matteo, T., Springel, V., Hernquist, L., 2005, Natur, 433, 604D
\bibitem[Di Matteo et al. 2008]{}Di Matteo, T., Colberg, J., Springel, V., Hernquist, L., Sijacki, D., 2008, ApJ, 676, 33D
\bibitem[Di Stefano 2007]{}Di Steffano, R., 2007, astro-ph/0712.3558
\bibitem[Dotti et al. 2007]{}Dotti, M., Colpi, M., Haardt, F., Lucio, M., 2007, MNRAS, 379, 956D   
\bibitem[Dotti et al. 2010]{}Dotti, M., Volonteri, M., Perego, A., Colpi, M., Ruszkowski, M., Haardt, F., 2010, MNRAS, 402, 682D 
\bibitem[Ebisuzaki et al. 1991]{}Ebisuzaki, T., Makino, J., Okumura, S.K., 1991, Natur, 354, 212E
\bibitem[Escala et al. 2005]{Escala}Escala, A, Larson, R.B., Coppi, P.S.,  
Mardones, D., 2005, ApJ, 630, 152
\bibitem[Fan 2005]{}Fan, X., 2005, gbha.conf, 75F
\bibitem[Fender et al. 2004]{}Fender, R.P., Belloni, T.M., Gallo, E., 2004, MNRAS, 355, 1105
\bibitem[Ferrarese \& Merritt 2000]{}Ferrarese, L., Merritt, D., 2000, ApJ, 539L, 9F
\bibitem[Ferrarese 2002]{}Ferrarese, L., 2002, ApJ, 578, 90F
\bibitem[Ferrarese et al. 2006]{}Ferrarese, L., et al. 2006, ApJ, 644L, 21F
\bibitem[Gebhardt et al. 2000]{}Gebhardt, K., et al. 2000, ApJ, 539L, 13G
\bibitem[Gebhardt et al. 2001]{}Gebhardt, K., et al. 2001, AJ, 122, 2469
\bibitem[Ghez et al. 2008]{}Ghez, A.M., et al , 2008, ApJ, 689, 1044G
\bibitem[Ghigna et al. 1998]{1998MNRAS.300..146G} Ghigna, S., Moore, B.,  
Governato, F., Lake, G., Quinn, T., Stadel, J., 1998, MNRAS, 300, 146 
\bibitem[Glover et al. 2010]{}Glover, S.C.O., 2010, astro-ph/1007.2763
\bibitem[Gnedin 2000]{}Gnedin, N, Y., 2000, ApJ, 542, 535G
\bibitem[Gonzalez et al. 2007a]{}Gonzalez, J.A., Hannam, M., Sperhake, U., Brugmann, B.,
Husa, S., 2007, PhRvL, 98w1101G   
\bibitem[Gonzalez et al. 2007b]{}Gonzalez, J.A., Sperhake, U., Brugmann, B., Hannam, M., 
Husa, S., 2007, PhRvL, 98i1101G   
\bibitem[Granato et al. 2001]{}Granato, G.L., Silva, L., Monaco, P., Panuzzo, P., 
Salucci, P., De Zotti, G., Danese, L., 2001, MNRAS, 324, 757G
\bibitem[Greenstein \& Matthews 1963]{}Greenstein, J.L., Matthews, T.A, 1963, AJ, 68S, 279G
\bibitem[Haehnelt \& Kauffmann 2000]{}Haehnelt, M.G., Kauffmann, G., 2000, MNRAS, 318L, 35H
\bibitem[Haehnelt et al. 1998]{}Haehnelt, M.G., Natarajan, P., Rees, M.J., 1998, MNRAS, 300, 817H
\bibitem[Heger et al. 2003]{Heger}Heger et al., 2003, ApJ, 591, 288H     
\bibitem[Heggie et al. 2007]{}Heggie, D.C., Hut, P., Mineshige, S., Makino, J., 
Baumgardt, H., 2007, PGASJ, 59L, 11H   
\bibitem[Hernquist 1989]{1989Natur.340..687H}Hernquist, L., 1989, Natur, 340, 687 
\bibitem[Herrmann et al. 2007]{}Herrmann, F., Hinder, I., Shoemaker, D., Laguna, P., 
Matzner, R.A., 2007, ApJ, 661, 430H   
\bibitem[Ho 2008]{}Ho, L., 2008, astro-ph/0803.2268  
\bibitem[Holley-Bockelmann \& Richstone 1999]{}Holley-Bockelmann, K., \&
Richstone, D.O., 1999, ApJ, 517, 92H
\bibitem[Holley-Bockelmann \& Sigurdsson 2006]{}Holley-Bockelmann, K., \&
Sigurdsson, S., 2006, astro-ph, 1520H   
\bibitem[Holley-Bockelmann et al. 2007]{} Holley-Bockelmann, K., Gultekin, K., Shoemaker, D., 
\& Yunes, N.\ 2007, ArXiv e-prints, 707, arXiv:0707.1334 
\bibitem[Holley-Bockelmann et al. 2010]{}Holley-Bockelmann, K., Micic, M., Sigurdsson, S., Rubbo, L., 2010, ApJ, 713, 1016 
\bibitem[Hopkins et al. 2005]{}Hopkins, P.F., Hernquist, L., Cox, T.J., Di Matteo, T.,
Martini, P., Robertson, B., Springel, V., 2005, ApJ, 630, 705H
\bibitem[Hopkins et al. 2007]{}Hopkins, P. F., Hernquist, L., Cox, T. J., Robertson, B., Krause, E., 
2007, ApJ, 669, 67H
\bibitem[Hopkins \& Hernquist 2009]{}Hopkins, P. F., Hernquist, L., 2009, ApJ, 698, 1550H
\bibitem[Hopkins et al. 2010]{}Hopkins, P. F., Murray, N., Quataert, E., Thompson, T. A., 2010, MNRAS, 401L, 19H
\bibitem[Hu et al. 2006]{}Hu, J., Shen, Y., Lou, Y., Zhang, S., 2006, ApJ, 365, 345   
\bibitem[Islam et al. 2003]{Islam}Islam, R. R., Taylor, J. E., Silk, J., 
2003, MNRAS, 340, 647I    
\bibitem[Islam et al. 2004]{Islam}Islam, R. R., Taylor, J. E., Silk, J., 
2004, MNRAS, 354, 427I
\bibitem[Kauffmann \& Haehnelt 2000]{}Kauffmann, G., Haehnelt, M., 2000, MNRAS, 311, 576K
\bibitem[Kazantzidis et al. 2005]{}Kazantzidis, S., Mayer, L., Colpi, M., 
Madau, P., Debattista, V.P., Wadsley, J., Stadel, J., Quinn, T., Moore, B., 
2005, ApJ, 623L, 67K     
\bibitem[Keres et al. 2005]{}Keres, D., Katz, N., Weinberg, D.H., Dave, R., 2005, MNRAS, 363, 2
\bibitem[Kesden et al. 2010]{}Kesden, M., Sperhake, U., Berti, E., 2010, ApJ, 715, 1006K
\bibitem[Komossa et al. 2008]{}Komossa, S., Zhou, H., Lu, H., 2008, ApJ, 678L, 81K 
\bibitem[Koppitz et al. 2007]{}Koppitz, M., Pollney, D., Reisswig, C., Rezzolla, L., 
Thornburg, J., Diener, P., Schnetter, E., 2007, gr.qc, 1163K   
\bibitem[Kormendy \& Richstone 1995]{}Kormendy, J., Richstone, D., 1995, ARA\&A, 33, 581K   
\bibitem[Koushiappas et al. 2004]{}Koushiappas, S.M., Bullock, J.S., Dekel, A., 2004, MNRAS, 354, 292K
\bibitem[Kravtsov et al. 2004]{}Kravtsov, A.V., Gnedin, O.Y., Klypin, A.A., 2004, ApJ, 609, 482K
\bibitem[Lehner \& Moreschi 2007]{}Lehner, L., Moreschi, O.M., 2007, PhRvD, 7614040L 
\bibitem[Loeb \& Rasio 1994]{}Loeb, A., Rasio, F.A., 1994, ApJ, 432, 52L
\bibitem[Lousto \& Zlochower 2009]{}Lousto, C.O., Zlochower, Y., 2009, PhRvD, 79f4018L
\bibitem[Maccarone 2004]{}Maccarone, T.J., 2004, MNRAS, 35, 1049M
\bibitem[Maccarone et al. 2005]{}Maccarone, T.J., Fender, R.P., Tzioumis, A.K., 2005, MNRAS, 356, L17 
\bibitem[Maccarone 2005]{}Maccarone, T.J., 2005, MNRAS, 360, L30
\bibitem[Mack et al. 2007]{}Mack, K.J., Ostriker, J.P., Ricotti, M., 2007, ApJ, 665, 1277M
\bibitem[Madau \& Rees 2001]{}Madau, P., Rees, M.J., 2001, ApJ, 551L, 27M
\bibitem[Makino 1997]{}Makino, J., 1997, ApJ, 478, 58M
\bibitem[Marconi \& Hunt 2003]{} Marconi, A., Hunt, L. K., 2003, ApJ, 589, L21
\bibitem[Marconi et al. 2004]{2004MNRAS.351..169M} Marconi, A., Risaliti,
G., Gilli, R., Hunt, L.K., Maiolino, R., Salvati, M., 2004, MNRAS, 351, 169 
\bibitem[Matthews \& Brighenti 2003]{}Matthews, W.G., Brighenti, F., 2003, ApJ, 599, 992M
\bibitem[Mayer et al. 2007]{}Mayer, L., Kazantzidis, S., Madau, P., Colpi, M., 
Quinn, T., Wadsley, J., 2007, Sci, 316, 1874M
\bibitem[McWilliams 2008]{}McWilliams, S.T., 2008, PhDT, 7M
\bibitem[Melia \& Falcke 2001]{}Melia, F., Falcke, H., 2001, ARA\&A, 39, 309M
\bibitem[Menou et al. 2001]{}Menou, K., Haiman, Z., Narayanan, V.K., 
2001, ApJ, 558, 535
\bibitem[Merloni 2004]{}Merloni, A., 2004, MNRAS, 353, 1035
\bibitem[Merritt et al. 2001]{}Merritt, D., Ferrarese, L., Joseph, C. L., 2001, Science, 293, 1116M
\bibitem[Merritt et al. 2004]{}Merritt, D., Milosavljevic, M., Favata, M., Hughes, S.A., 
Holz, D.E., 2004, ApJ, 607L, 9M  
\bibitem[Mihos \& Hernquist 1994]{}Mihos, J.C., Hernquist, L., 1994, ApJ, 425L, 13M
\bibitem[Miller \& Colbert 2004]{}Miller, M.C., Colbert, E.J.M., 2004, IJMPD, 13, 1M
\bibitem[Milosavljevic \$ Merritt 2001]{}Milosavljevic, M., Merritt, D., 2001, ApJ, 563, 34M
\bibitem[Milosavljevic \$ Merritt 2003]{}Milosavljevic, M., Merritt, D., 2003, ApJ, 596, 860M
\bibitem[Micic et al. 2006]{Micic}Micic, M., Abel, T., Sigurdsson, S., 2006, MNRAS, 372, 1540M 
\bibitem[Micic et al. 2007]{}Micic, M., Holley-Bockelmann, K., Sigurdsson, S., Abel, T., 
2007, MNRAS, 380, 1533M  
\bibitem[Monaco et al. 2000]{}Monaco, P., Salucci, P., Danese, L., 2000, MNRAS, 311, 279M
\bibitem[Navarro, Frenk \& White 1996]{}Navarro, J.F., Frenk, C.S., White, S.D.M., 
1995, ApJ, 462, 563N   
\bibitem[Nayakshin et al. 2009]{}Nayakshin, S., Wilkinson, M. I., King, A., 2009, MNRAS, 398, 54N
\bibitem[Ostriker et al. 2010]{}Ostriker, J.P., Choi, E., Ciotti, L., Novak, G.S., Proga, D., 2010, ArXiv e-prints, 1004, arXiv:1004.2923 
\bibitem[O'Sullivan et al. 2001]{}O'Sullivan, E., Forbes, D.A., Pnman, T.J., 2001, MNRAS, 328, 461O
\bibitem[Pedersen et al. 2006]{}Pedersen, K., Rasmussen, J., Sommer-Larsen, J., Toft, S., Benson, A.J., Bower, R.G., 2006, NewA, 11, 465P
\bibitem[Portegies Zwart et al. 2004]{}Portegies Zwart, S.F., Baumgardt, H., Hut, P., 
Makino, J., McMillan, S.L.W., 2004, Natur, 428, 724P
\bibitem[Quataert \& Gruzinov 2000]{}Quataert, E., Gruzinov, A., 2000, ApJ, 545, 842Q
\bibitem[Quinlan 1996]{}Quinlan, G.D., 1996, NewA, 1, 35Q
\bibitem[Rees 1984]{}Rees, M.J., 1984, ARA\&A, 22, 471R
\bibitem[Richstone 1976]{} Richstone, D.O., 1976, ApJ, 204, 642 
\bibitem[Robertson et al. 2006]{}Robertson, B., Hernquist, L., Cox, T.J., Di Matteo, T., 
Hopkins, P.F., Martini, P., Springel, V., 2006, ApJ, 641, 90R
\bibitem[Sales et al. 2007]{2007MNRAS.379.1464S} Sales, L.V., Navarro,   
J.F., Abadi, M.G., Steinmetz, M., 2007, MNRAS, 379, 1464 
\bibitem[Schneider et al. 2002]{Schneider}Schneider, R., Ferrara, A., 
Natarajan, P., Omukai, K., 2002, ApJ, 571, 30S    
\bibitem[Schnittman 2007]{}Schnittman, J,D., 2007, ApJ, 667L, 133S
\bibitem[Schnittman \& Buonanno 2007]{}Schnittman, J.D., Buonanno, A., 2007, ApJ, 662L, 63S  
\bibitem[Sesana 2007]{}Sesana, A., 2007, MNRAS, 382L, 6S
\bibitem[Shakura \& Syunyaev 1973]{}Shakura, N.I., Syunyaev, R.A., 1973, A\&A, 24, 337S
\bibitem[Shankar et al. 2006]{}Shankar, F., Lapi, A., Salucci, P., De Zotti, G., Danese, L., 
2006, ApJ, 643, 14S
\bibitem[Sigurdsson 2003]{}Sigurdsson, S., 2003, CQGra, 20S, 45S 
\bibitem[Sijacki et al. 2007]{}Sijacki, D., Springel, V., di Matteo, T., Hernquist, L., 2007, MNRAS, 380, 877S
\bibitem[Sijacki et al. 2009]{}Sijacki, D., Springel, V., Haehnelt, M.G., 2009, MNRAS, 400, 100S
\bibitem[Silk \& Rees 1998]{}Silk, J., Rees, M.J., 1998, A\&A, 331L, 1S
\bibitem[Sinha \& Holley-Bockelmann 2010]{}Sinha, M., Holley-Bockelmann, K., 2010, MNRAS, 405L, 31S
\bibitem[Soltan 1982]{}Soltan A., 1982, MNRAS, 200, 115S
\bibitem[Somerville et al. 2008]{}Somerville, R., Hopkins, P.F., Cox, T.J., Robertson, B.E., Hernquist, L., 2008, MNRAS, 391, 481S
\bibitem[Sperhake 2009]{}Sperhake, U., 2009, LNP, 769, 125S
\bibitem[Taffoni et al. 2003]{}Taffoni, G., Mayer, L., Colpi, M., Governato, F.,
2003, MNRAS, 341, 434T    
\bibitem[Tremaine et al. 2002]{}Tremaine, S., et al. 2002, ApJ, 574, 740T
\bibitem[Trenti \& Stiavelli 2009]{}Trenti, M., Stiavelli, M., 2009, ApJ, 694, 879T
\bibitem[Trenti et al. 2009]{}Trenti, M., Stiavelli, M., Michael S. J., 2009, ApJ, 700, 1672L
\bibitem[Valluri et al. 2005]{}Valluri, M., Ferrarese, L., Merritt, D., Joseph, C. L. 2005, ApJ, 628, 137
\bibitem[van der Marel 2004]{}van der Marel, R.P., 2004, cbhg.symp, 37V
\bibitem[Volonteri et al. 2003]{Volonteri}Volonteri, M., Haardt, F., 
Madau, P., 2003, ApJ, 582, 559
\bibitem[Volonteri 2007]{}Volonteri, M., 2007, ApJ, 663L, 5V    
\bibitem[Volonteri et al. 2010]{}Volonteri, M., Gultekin, K., Dotti, M., 2010, astro-ph/1001.1743 
\bibitem[Weinberg 1989]{1989MNRAS.239..549W}Weinberg, M.D., 1989, MNRAS, 239, 549 
\bibitem[Wise  \&  Abel 2005]{Wise}Wise, J., H., Abel, T., 2005, ApJ, 629, 615W   
\bibitem[Wyithe \& Loeb 2005]{}Wyithe, J.S.B., Loeb, A., 2005, ApJ, 634, 910W


\end{thebibliography}
\end{document}